\documentclass{jcgt}
\usepackage[dvipsnames]{xcolor}

\setciteauthor{Jakub Bok\v{s}ansk\'y and Daniel Meister}
\setcitetitle{Neural Visibility Cache for Real-Time Light Sampling}

\accepted{2025-06-04}{2025-07-02}{2025-08-18}{Brent Burley}{14}{2}{1}{19}{2025}
\seturl{http://jcgt.org/published/0014/02/01/}

\usepackage[T1]{fontenc}
\usepackage{dfadobe}  
\usepackage{listings}
\usepackage{multirow}
\usepackage{lineno}
\usepackage{subfig}
\usepackage{amsmath}
\usepackage{amssymb}
\usepackage{bm}
\usepackage{tikz}
\usepackage[absolute,overlay]{textpos}

\newcommand{\includeimgcrop}[1]{\includegraphics[height=0.065\textheight, trim=75pt 0 75pt 0, clip]{#1}}
\newcommand{\includeimgcroplarge}[1]{\includegraphics[height=0.1\textheight, trim=0 0 0 0, clip]{#1}}

\begin{document}

\title{Neural Visibility Cache for\\ Real-Time Light Sampling}

\author
{
    Jakub Bok\v{s}ansk\'y~\href{https://orcid.org/0000-0003-0087-2645}{\includegraphics[width=8pt]{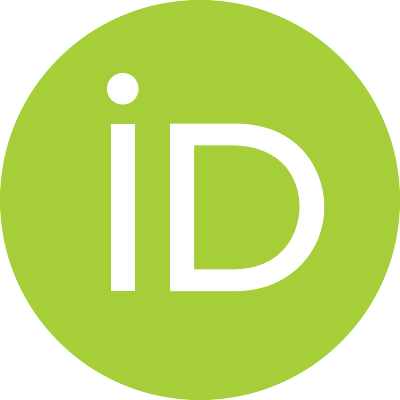}}\\Advanced Micro Devices, Inc.
\and
    Daniel Meister~\href{https://orcid.org/0000-0002-3149-1442}{\includegraphics[width=8pt]{ORCIDlogo}}\\Advanced Micro Devices, Inc.
}

\teaser{
\vspace*{-6pt}\centering
\begin{minipage}{0.5\textwidth}
    \begin{tikzpicture}
        \node[anchor=south west, inner sep=0] (image) at (0,0) {\includegraphics[width=\linewidth]{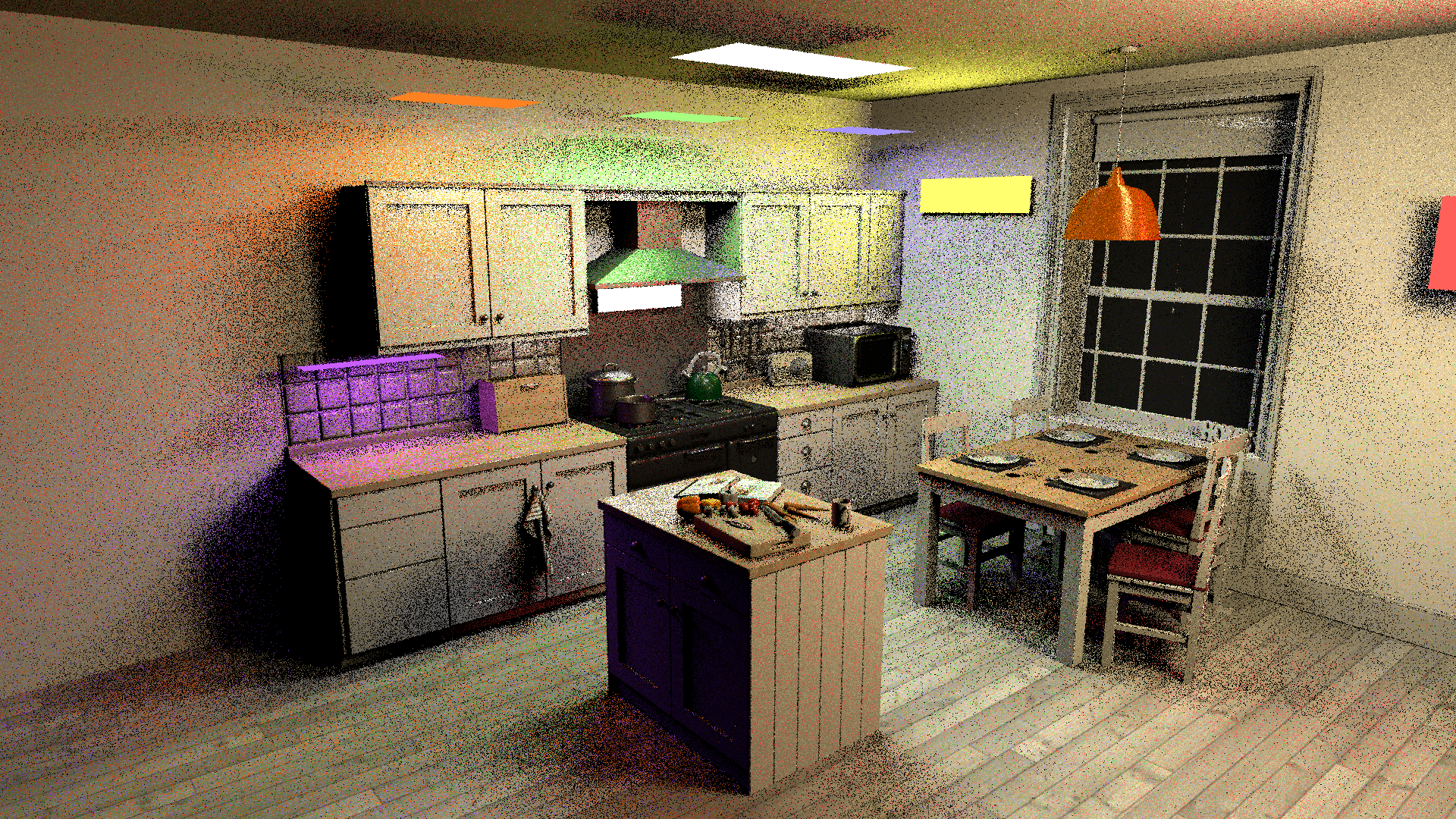}};
        \node[anchor=south east, xshift=-2mm, yshift=2mm, fill=white, opacity=0.7, text opacity=1, font=\scriptsize] 
            at (image.south east) {FLIP 0.361};
        \node[anchor=north west, xshift=2mm, yshift=-2mm, fill=white, opacity=0.7, text opacity=1, font=\scriptsize] 
            at (image.north west) {ReSTIR};
    \end{tikzpicture}
\end{minipage}%
\begin{minipage}{0.5\textwidth}
    \begin{tikzpicture}
        \node[anchor=south west, inner sep=0] (image) at (0,0) {\includegraphics[width=\linewidth]{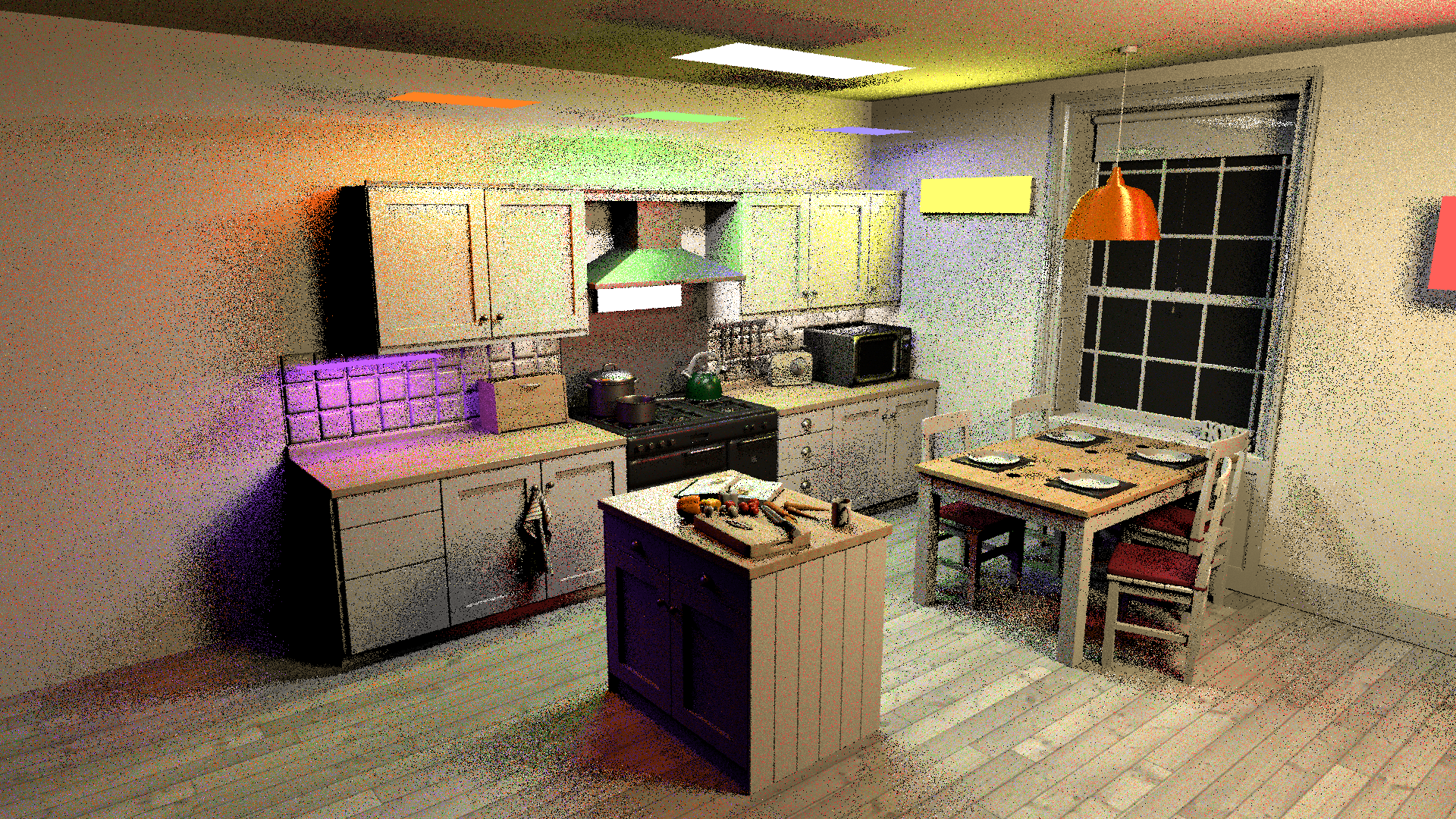}};
        \node[anchor=south east, xshift=-2mm, yshift=2mm, fill=white, opacity=0.7, text opacity=1, font=\scriptsize] 
            at (image.south east) {FLIP 0.298};
        \node[anchor=north west, xshift=2mm, yshift=-2mm, fill=white, opacity=0.7, text opacity=1, font=\scriptsize] 
            at (image.north west) {Neural Light Sampling (ours)};
    \end{tikzpicture}
\end{minipage}

\begin{minipage}{0.5\textwidth}
    \begin{tikzpicture}
        \node[anchor=south west, inner sep=0] (image) at (0,0) {\includegraphics[width=\linewidth]{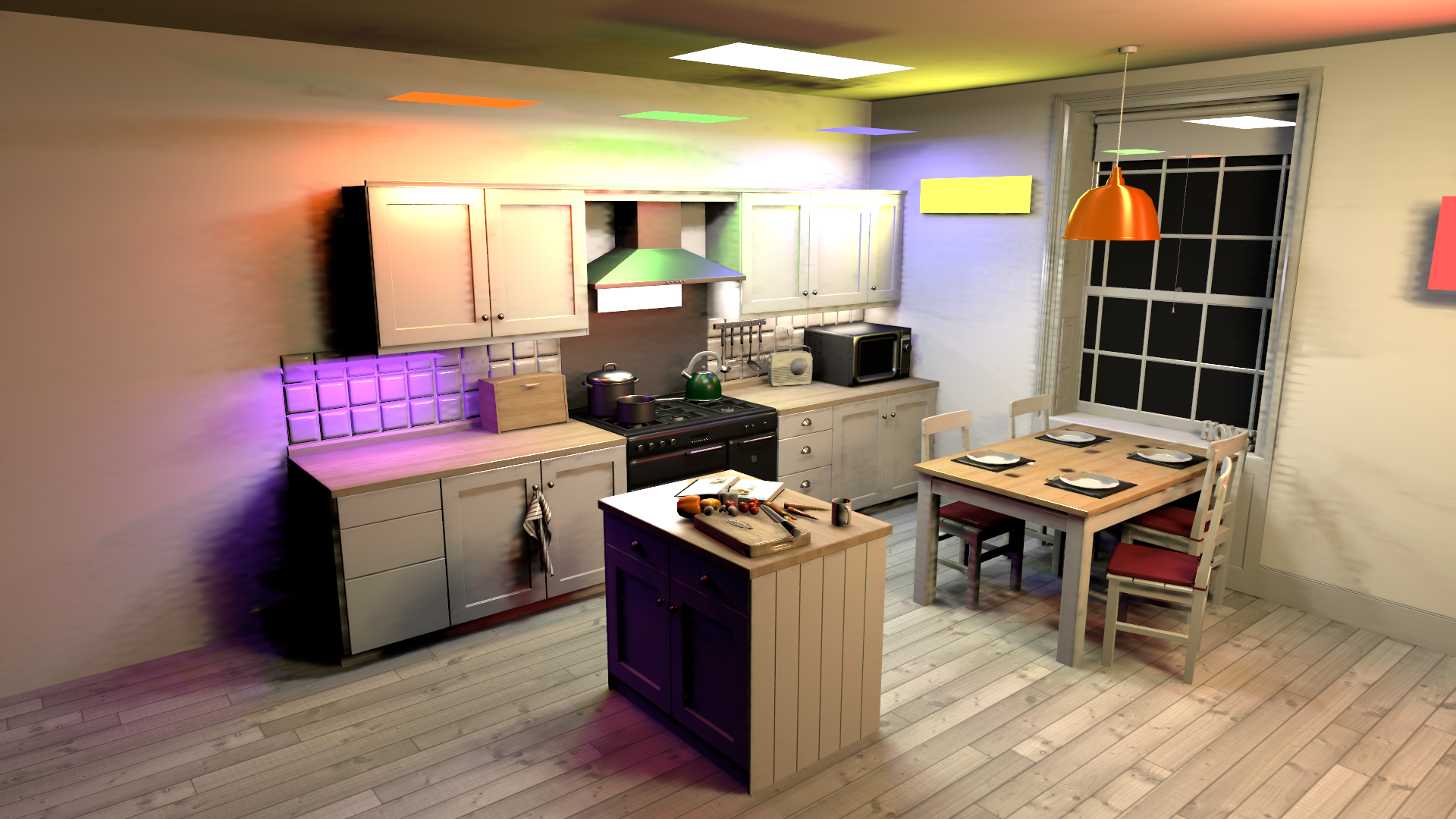}};
        \node[anchor=south east, xshift=-2mm, yshift=2mm, fill=white, opacity=0.7, text opacity=1, font=\scriptsize] 
            at (image.south east) {FLIP 0.08};
        \node[anchor=north west, xshift=2mm, yshift=-2mm, fill=white, opacity=0.7, text opacity=1, font=\scriptsize] 
            at (image.north west) {Neural DI (ours)};
    \end{tikzpicture}
\end{minipage}%
\begin{minipage}{0.5\textwidth}
    \begin{tikzpicture}
        \node[anchor=south west, inner sep=0] (image) at (0,0) {\includegraphics[width=\linewidth]{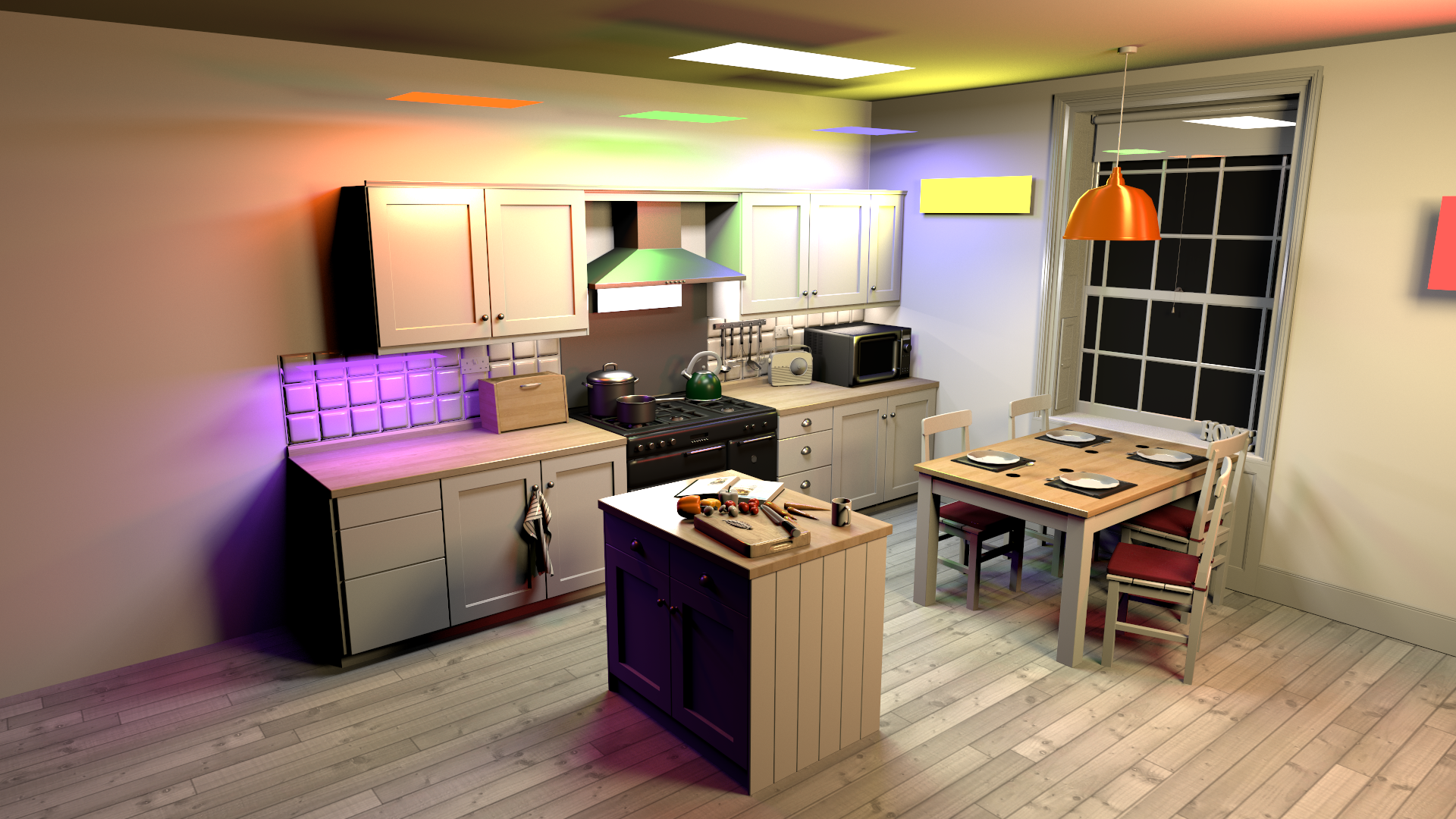}};
        \node[anchor=south west, inner sep=0] (image) at (0,0) {\includegraphics[width=\linewidth]{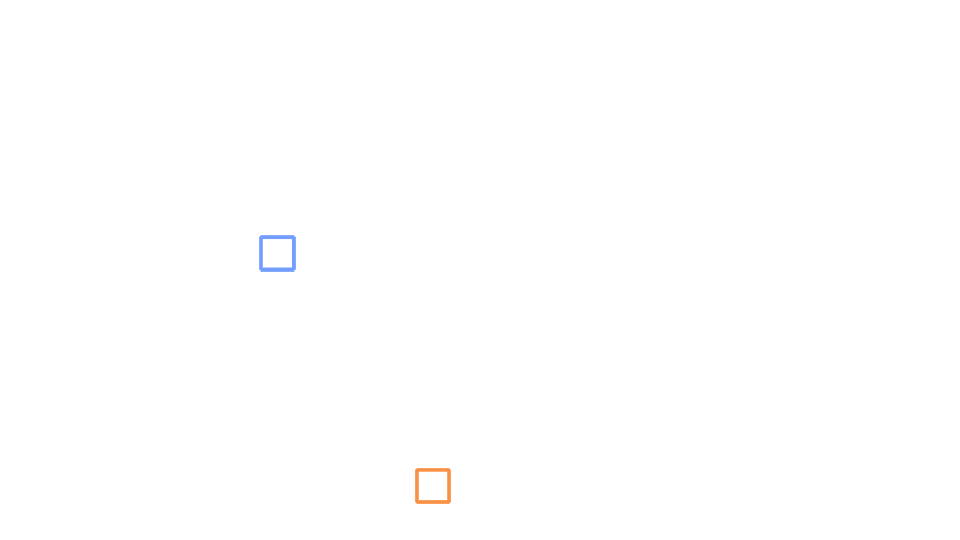}};
        \node[anchor=north west, xshift=2mm, yshift=-2mm, fill=white, opacity=0.7, text opacity=1, font=\scriptsize] 
            at (image.north west) {Ground Truth};
    \end{tikzpicture}
\end{minipage}%

\begin{minipage}{0.5\textwidth}
    \begin{tikzpicture}
        \node[anchor=south west, inner sep=0] (image) at (0,0) {\includegraphics[width=\linewidth]{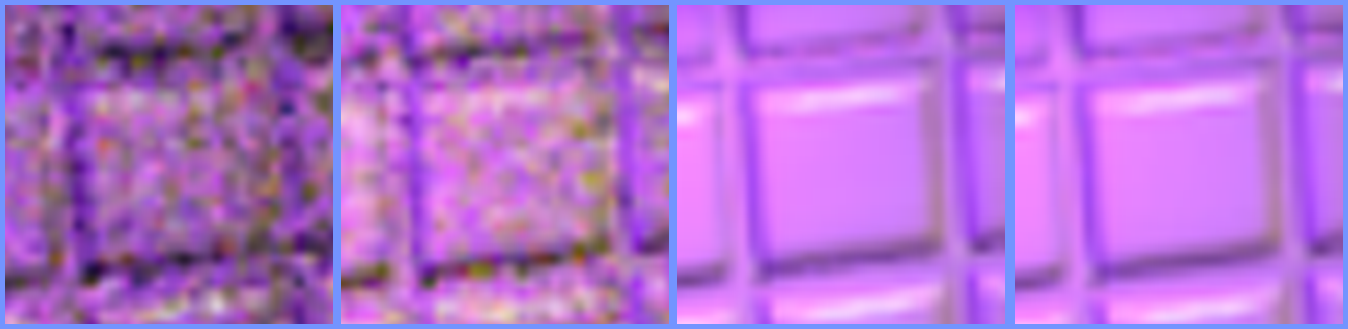}};
    \end{tikzpicture}
\end{minipage}%
\begin{minipage}{0.5\textwidth}
    \begin{tikzpicture}
        \node[anchor=south west, inner sep=0] (image) at (0,0) {\includegraphics[width=\linewidth]{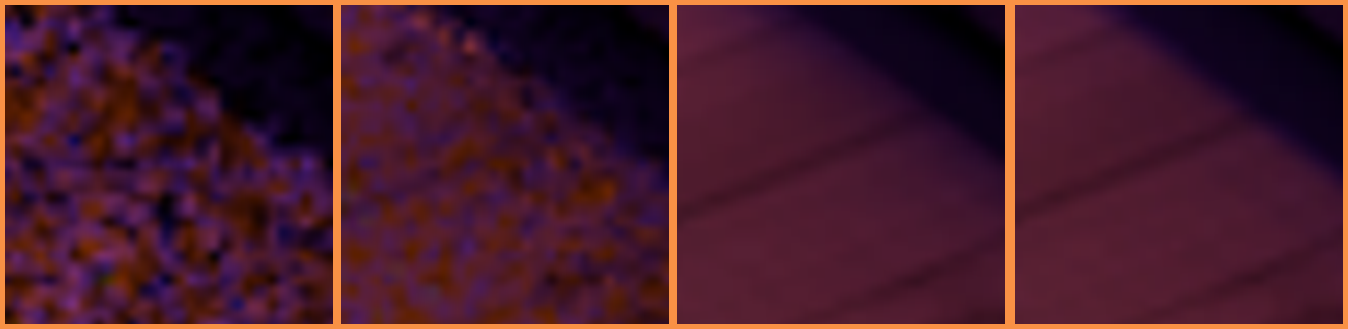}};
    \end{tikzpicture}
\end{minipage}%

\begin{minipage}{0.125\textwidth}
    \centering
{\scriptsize ReSTIR }
\end{minipage}%
\begin{minipage}{0.125\textwidth}
    \centering
{\scriptsize NLS }
\end{minipage}%
\begin{minipage}{0.125\textwidth}
    \centering
{\scriptsize Neural DI }
\end{minipage}%
\begin{minipage}{0.125\textwidth}
    \centering
{\scriptsize Ground Truth }
\end{minipage}%
\begin{minipage}{0.125\textwidth}
    \centering
{\scriptsize ReSTIR }
\end{minipage}%
\begin{minipage}{0.125\textwidth}
    \centering
{\scriptsize NLS }
\end{minipage}%
\begin{minipage}{0.125\textwidth}
    \centering
{\scriptsize Neural DI }
\end{minipage}%
\begin{minipage}{0.125\textwidth}
    \centering
{\scriptsize Ground Truth }
\end{minipage}%

\caption{Rendering of the kitchen scene with 32 lights using screen-space ReSTIR (top left) at 3.61~ms per frame, compared to our neural light sampling method (NLS) (top right) running at 3.95~ms, our neural direct illumination (Neural DI) (bottom left) at 4.41~ms, and ground truth (bottom right). Compared to screen-space ReSTIR, our neural light sampling produces less noise (especially in occluded areas), reducing the FLIP error metric~\cite{Andersson2021b}. Additionally, we can use the neural network to compute approximate direct lighting without the necessity to cast shadow rays, except for the rays needed for neural network training.}
\label{fig:teaser}
}

\maketitle
\thispagestyle{firstpagestyle}

\vspace*{-6pt}
\begin{abstract}
\small
Direct illumination with many lights is an inherent component of physically-based rendering that remains challenging, especially in real-time scenarios. We propose an online-trained neural cache that stores visibility between lights and 3D positions. We feed light visibility to weighted reservoir sampling (WRS)~\cite{chao1982general, Wyman2021WRS} to sample a light source. The cache is implemented as a fully-fused multilayer perceptron (MLP)~\cite{muller2021real} with multi-resolution hash-grid encoding~\cite{muller2022instant}, enabling online training and efficient inference on modern GPUs in real-time frame rates. The cache can be seamlessly integrated into existing rendering frameworks and can be used in combination with other real-time techniques such as spatiotemporal reservoir sampling (ReSTIR)~\cite{bitterli20spatiotemporal}.
\end{abstract}

\section{Introduction}
Direct illumination has a significant impact on both the quality of rendered images and the rendering performance. The reflected radiance due to direct illumination at point $\mathbf{x}$ in direction $\mathbf{\omega}_o$ can be described as an integral over all light emitting surfaces~$A$:
\begin{equation}
L(\mathbf{x},\mathbf{\omega}_o) =\int \limits_{A}  f_r(\mathbf{x}, \mathbf{\omega}_{\mathbf{x} \xrightarrow{} \mathbf{y}}, \mathbf{\omega}_o) L_e(\mathbf{x}, \mathbf{\omega}_{\mathbf{y} \xrightarrow{} \mathbf{x}})G(\mathbf{x}, \mathbf{y}) V(\mathbf{x}, \mathbf{y}) \mathrm{d}A(\mathbf{y}),
\label{eq:reflection}
\end{equation}
where $f_r$ is the bidirectional reflectance function (BRDF), $L_e$ is the emitted radiance, $\mathbf{\omega}_{\mathbf{x} \xrightarrow{} \mathbf{y}}$ is a unit direction pointing from point $\mathbf{x}$ to $\mathbf{y}$, $G$ is the geometry term including cosine terms and the squared distance, and $V$ is the visibility function indicating binary visibility between two points.

We solve Equation~\eqref{eq:reflection} by means of Monte Carlo integration as there is no general analytic solution. As with any method based on Monte Carlo integration, the challenging part is to find a probability density function that closely matches the desired distribution. To tackle this problem, we train a neural network to provide estimates of visibility between light sources and 3D positions that we use to guide the sampling process. 

We utilize weighted reservoir sampling (WRS)~\cite{chao1982general, Wyman2021WRS} to sample a light source based on the light visibility estimated by the neural network and BRDF contribution to the shaded point, providing an unbiased sampling mechanism (see Section~\ref{sec:OurMethod}). 
The network architecture is based on a fully-fused multilayer perceptron (MLP)~\cite{muller2021real}, which allows for efficient online training and inference on contemporary GPUs in real-time frame rates. We employ multi-resolution hash-grid encoding~\cite{muller2022instant} to learn high-frequency details in a lower-dimensional space. The proposed method can be easily integrated into existing real-time rendering pipelines. For instance, it can be used in next event estimation for reflected bounces in path tracing, or for spatiotemporal reservoir sampling (ReSTIR)~\cite{bitterli20spatiotemporal} to sample initial candidates or to recover after abrupt visibility changes that might otherwise cause significant noise. Our neural representation of visibility works with either individual lights directly or their clusters, to support scenes with an arbitrary number of lights.

\section{Previous Work}
Global illumination algorithms are notoriously known for their high computational demands. Therefore, a vast body of acceleration techniques has been proposed. Among these, various caching strategies play an important role, including irradiance caching~\cite{ward1994adaptive}, which was later generalized to radiance caching~\cite{Krivanek2005RCE}. These approaches are generally biased due to interpolation. To avoid bias, the idea is to reuse cached information only to guide the sampling process to better match the target distribution. Thus, numerous path guiding algorithms have been proposed, employing algorithms such as hierarchical data structures~\cite{mueller2017practical, yusuke24}, online-trained parametric mixture models~\cite{Vorba2014OLP}, neural networks~\cite{muller2019neural}, or a combination of these~\cite{huang2024}.

With the advent of many-light rendering~\cite{Dachsbacher2014ManyLights}, the problem of global illumination reduces to direct lighting with a large number of virtual point lights, necessitating efficient light sampling techniques. Traditional solutions are either based on arranging light sources into a hierarchical data structure~\cite{Walter2005, Estevez2018, moreau2019dynamic}, sampling the light transport matrix~\cite{hasan2007}, visibility-aware reinforcement learning~\cite{Pantaleoni2019}, or Bayesian inference~\cite{Vevoda2018BOR}. \citet{guo_nee} presented a method for caching visibility between voxels, significantly reducing the number of precise visibility tests required. Most of the aforementioned methods are designed for offline rendering; the overhead is prohibitively expensive for real-time applications. \citet{li2024cache} presented an online-trained hierarchical light cache for sampling a very large number of lights in an unbiased manner in a production renderer.

With hardware acceleration of deep learning and ray tracing, physically-based rendering and neural-based approaches have become more compelling for real-time applications. Spatiotemporal reservoir sampling (ReSTIR)~\cite{bitterli20spatiotemporal} became the de facto standard for sampling direct lighting in real-time ray tracing, forgoing building any complex data structures and exploiting spatial and temporal correlation to efficiently process a very large number of lights. Several neural-based approaches have been recently proposed for real-time scenarios: a neural radiance cache~\cite{muller2021real}, neural shadow mapping~\cite{datta2022neural}, a neural light grid for precomputed indirect lighting~\cite{iwanicki2024lightgrid}, and a neural-based rendering framework employing an attention mechanism to solve the many-light problem~\cite{ren2024}. Concurrently to our work, \citet{dereviannykh2025neural} suggest to use neural incidence radiance caching in combination with two-level Monte Carlo integration to achieve unbiased estimates. The authors also proposed to cache visibility of the environment map lighting as a special case. A neural importance sampling of many lights~\cite{figueiredo2025neural} combines a light hierarchy with a neural network to predict light selection distribution directly, based on reflected radiance.

\section{Neural Visibility Cache}
\label{sec:OurMethod}

\subsection{Algorithm Outline}
\label{sec:Outline}

\begin{figure}
\centering
\includegraphics[width=0.9\textwidth]{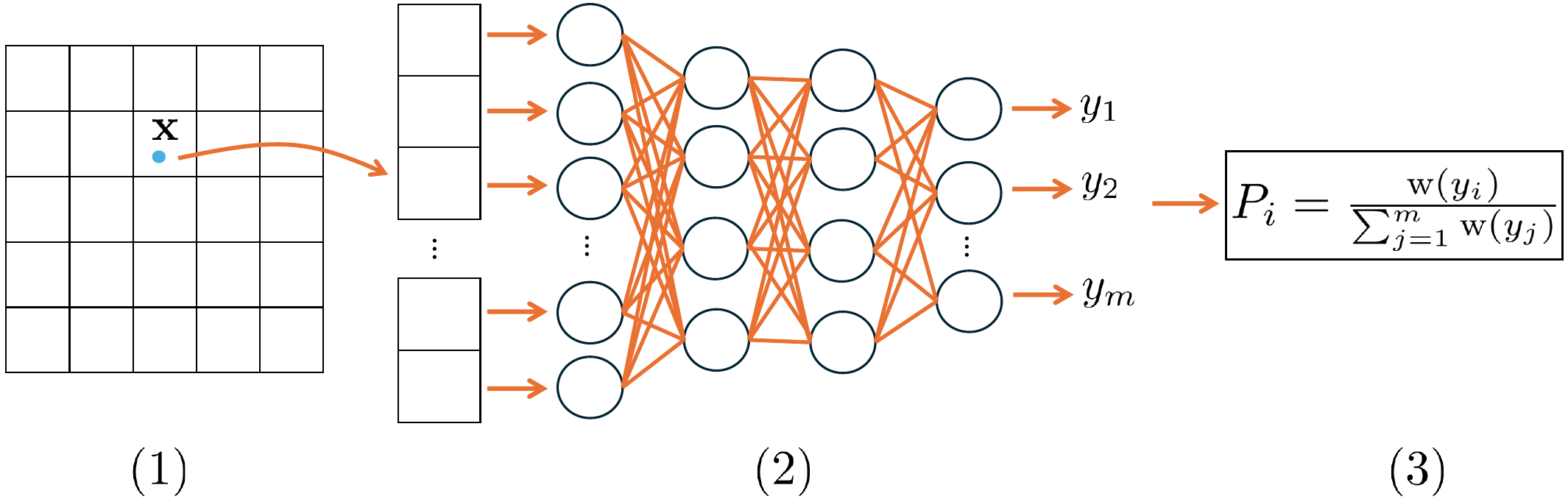}
\caption{Overview of our method. We use a multi-resolution hash-grid encoding~\cite{muller2022instant} to encode a 3D position (1), which is fed to a neural network (2). The output of the network is plugged into weighted reservoir sampling (WRS)~\cite{chao1982general, Wyman2021WRS} (3).}
\label{fig:overview}
\end{figure}

\begin{figure*}
\centering
\begin{minipage}{0.33\textwidth}
    \begin{tikzpicture}
        \node[anchor=south west, inner sep=0] (image) at (0,0) {\includegraphics[width=\linewidth]{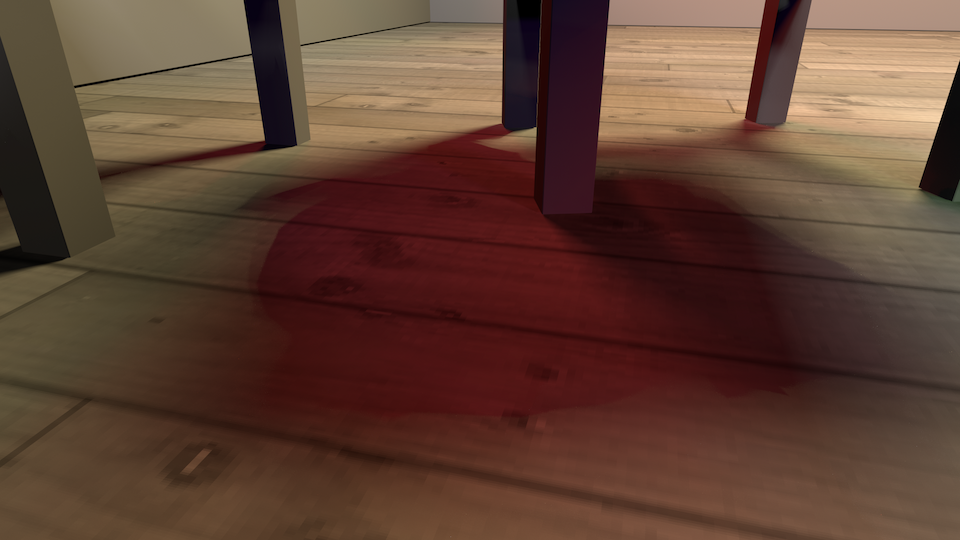}};
        \node[anchor=north west, xshift=1mm, yshift=-1mm, fill=white, opacity=0.7, text opacity=1, font=\tiny] 
            at (image.north west) {Biased NLS (Leaky ReLU)};
    \end{tikzpicture}
\end{minipage}%
\begin{minipage}{0.33\textwidth}
    \begin{tikzpicture}
        \node[anchor=south west, inner sep=0] (image) at (0,0) {\includegraphics[width=\linewidth]{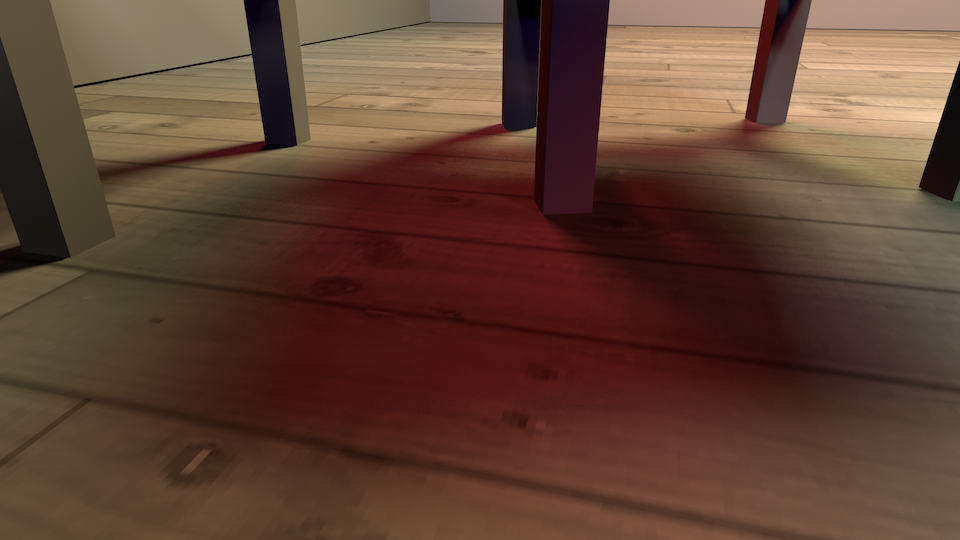}};
        \node[anchor=north west, xshift=1mm, yshift=-1mm, fill=white, opacity=0.7, text opacity=1, font=\tiny] 
            at (image.north west) {Unbiased NLS (Leaky ReLU)};
    \end{tikzpicture}
\end{minipage}%
\begin{minipage}{0.33\textwidth}
    \begin{tikzpicture}
        \node[anchor=south west, inner sep=0] (image) at (0,0) {\includegraphics[width=\linewidth]{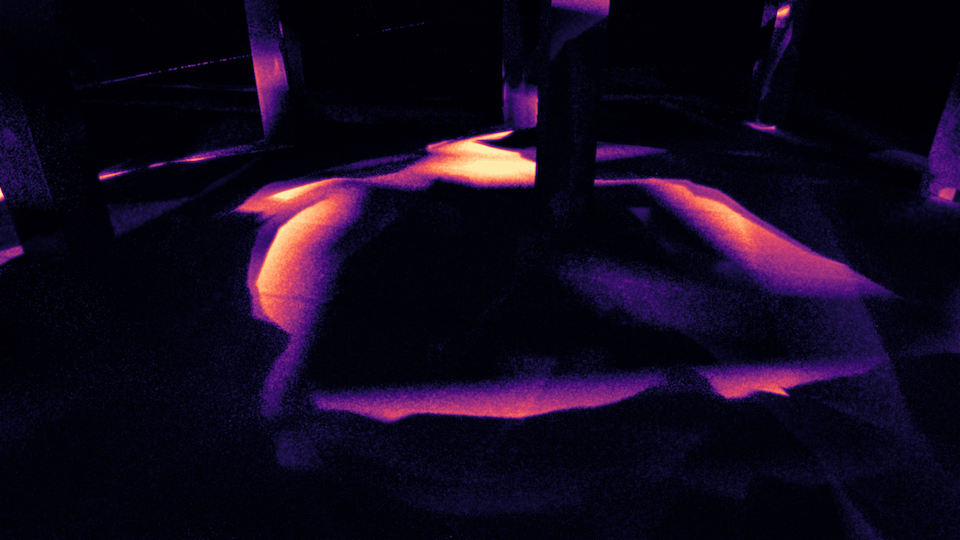}};
        \node[anchor=south east, xshift=-1mm, yshift=1mm, fill=white, opacity=0.7, text opacity=1, font=\tiny] 
            at (image.south east) {FLIP 0.007};
    \end{tikzpicture}
\end{minipage}%

\begin{minipage}{0.33\textwidth}
    \begin{tikzpicture}
        \node[anchor=south west, inner sep=0] (image) at (0,0) {\includegraphics[width=\linewidth]{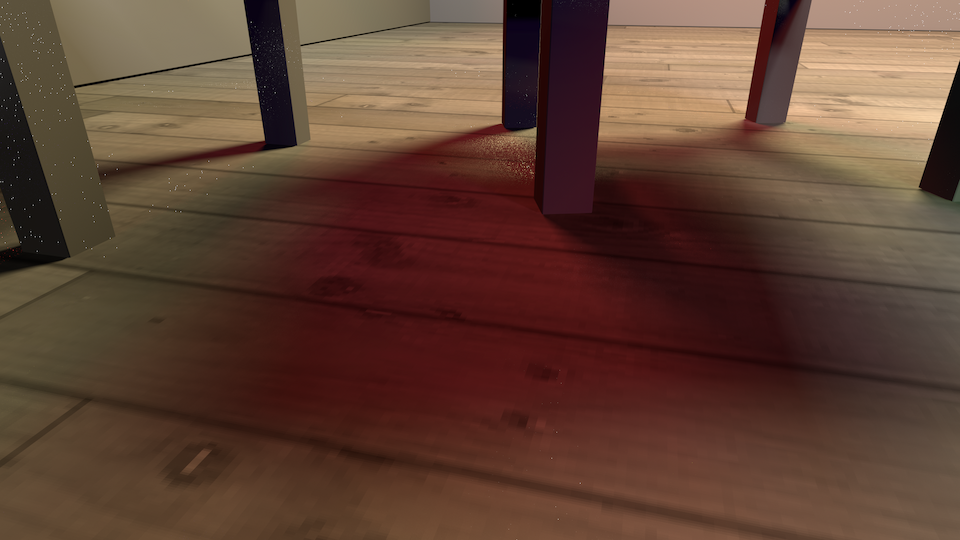}};

        \node[anchor=north west, xshift=1mm, yshift=-1mm, fill=white, opacity=0.7, text opacity=1, font=\tiny] 
            at (image.north west) {Biased NLS (Sigmoid)};
    \end{tikzpicture}
\end{minipage}%
\begin{minipage}{0.33\textwidth}
    \begin{tikzpicture}
        \node[anchor=south west, inner sep=0] (image) at (0,0) {\includegraphics[width=\linewidth]{44_unbiased.png}};
        \node[anchor=north west, xshift=1mm, yshift=-1mm, fill=white, opacity=0.7, text opacity=1, font=\tiny] 
            at (image.north west) {Unbiased NLS (Sigmoid)};
    \end{tikzpicture}
\end{minipage}%
\begin{minipage}{0.33\textwidth}
    \begin{tikzpicture}
        \node[anchor=south west, inner sep=0] (image) at (0,0) {\includegraphics[width=\linewidth]{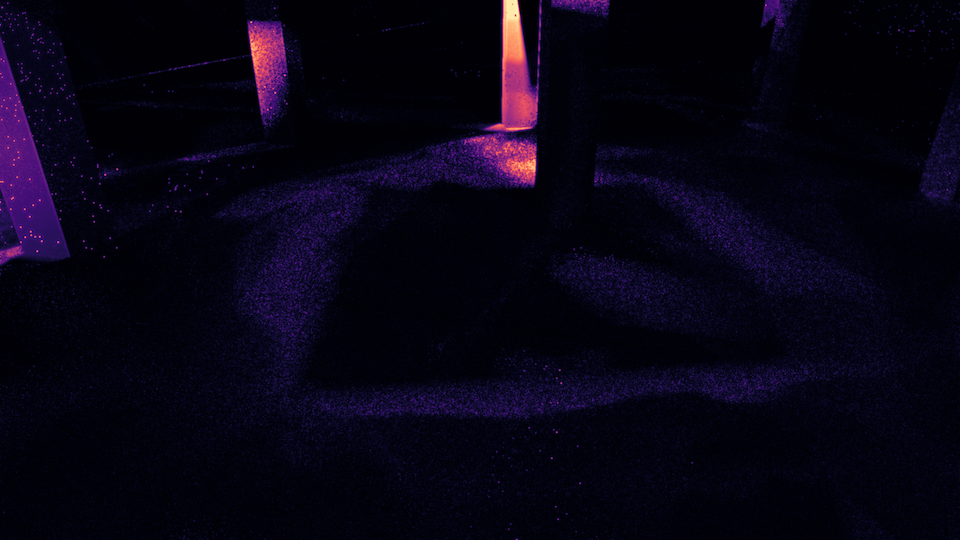}};
        \node[anchor=south east, xshift=-1mm, yshift=1mm, fill=white, opacity=0.7, text opacity=1, font=\tiny] 
            at (image.south east) {FLIP 0.004};
    \end{tikzpicture}
\end{minipage}%
\caption{\label{fig:bias}
 Our method can produce a biased result (left column) that typically exhibits as a hard boundary around heavily shadowed areas or as fireflies. The top row shows neural light sampling (NLS) with leaky rectified linear unit (ReLU) as the activation function for the output layer, which accentuates hard shadow boundary artifacts. The bottom row uses sigmoid for output activation, which has more fireflies instead. Clamping the visibility to 0.001 (center column) yields an unbiased result at the cost of slightly higher variance, alleviating these artifacts and  converging to the ground truth. Images have been accumulated for 16K frames.}
\end{figure*}

In this section, we describe our \emph{neural visibility cache} (NVC) and how to use it for light sampling. We train a multilayer perceptron~(MLP) to predict (nonbinary) visibility between any point and any light source (or a cluster of light sources) in the scene. The nonbinary visibility accounts for soft shadows cast by area lights and semitransparent surfaces. Given a 3D position in the scene, we first use the multi-resolution hash-grid encoding~\cite{muller2022instant} to encode the position, which we subsequently feed into the MLP, which outputs an estimated visibility for each light source; each neuron of the output layer corresponds to a single light. Finally, we use weighted reservoir sampling (WRS)~\cite{chao1982general, Wyman2021WRS} to sample a light source using the visibility estimates provided by the MLP (see Figure~\ref{fig:overview}). To ensure our method remains unbiased, we clamp zero and possibly negative values to a small positive constant (see Figure~\ref{fig:bias}). This introduces a slight amount of noise while avoiding bias.

We approximate the product of BRDF and the cosine term (see Equation~\eqref{eq:reflection}) by linearly transformed cosines (LTC)~\cite{heitz2016real}; we then multiply this by the light radiance and visibility predicted by the neural network to calculate the weights of the lights contributing to the shaded points for WRS. Thus, the probabilities of selecting each sample account for both visibility and BRDF, reducing noise in shadowed areas and penumbras as well. In contrast to methods based on estimating radiance~\cite{figueiredo2025neural}, we only estimate visibility and calculate exact reflected radiance analytically. This leads to faster training of the network, requiring less training iterations to converge.

Our algorithm can be used as a standalone or as a generator of initial candidates for ReSTIR (see Section~\ref{sec:restir_with_NVC}). This increases convergence speed and reduces noise in disoccluded pixels where ReSTIR can struggle to find meaningful initial light samples. ReSTIR implementations typically cast a shadow ray for a selected initial candidate and invalidate it if it is occluded. When generating initial candidates using our method, this is unnecessary as we already take visibility into account for all candidates.

To solve a many-lights problem, we employ a clustered approach, where each output neuron represents the average visibility of a light cluster, instead of an individual light source. This way, we can support an arbitrary number of lights, sorted into a fixed number of clusters. Sampling then becomes a two-step process, where a cluster of lights is selected first, and then a light sample within the cluster.

\begin{figure}
\centering
\begin{minipage}{0.485\textwidth}
    \begin{tikzpicture}
        \node[anchor=south west, inner sep=0] (image) at (0,0) {\includegraphics[width=\linewidth]{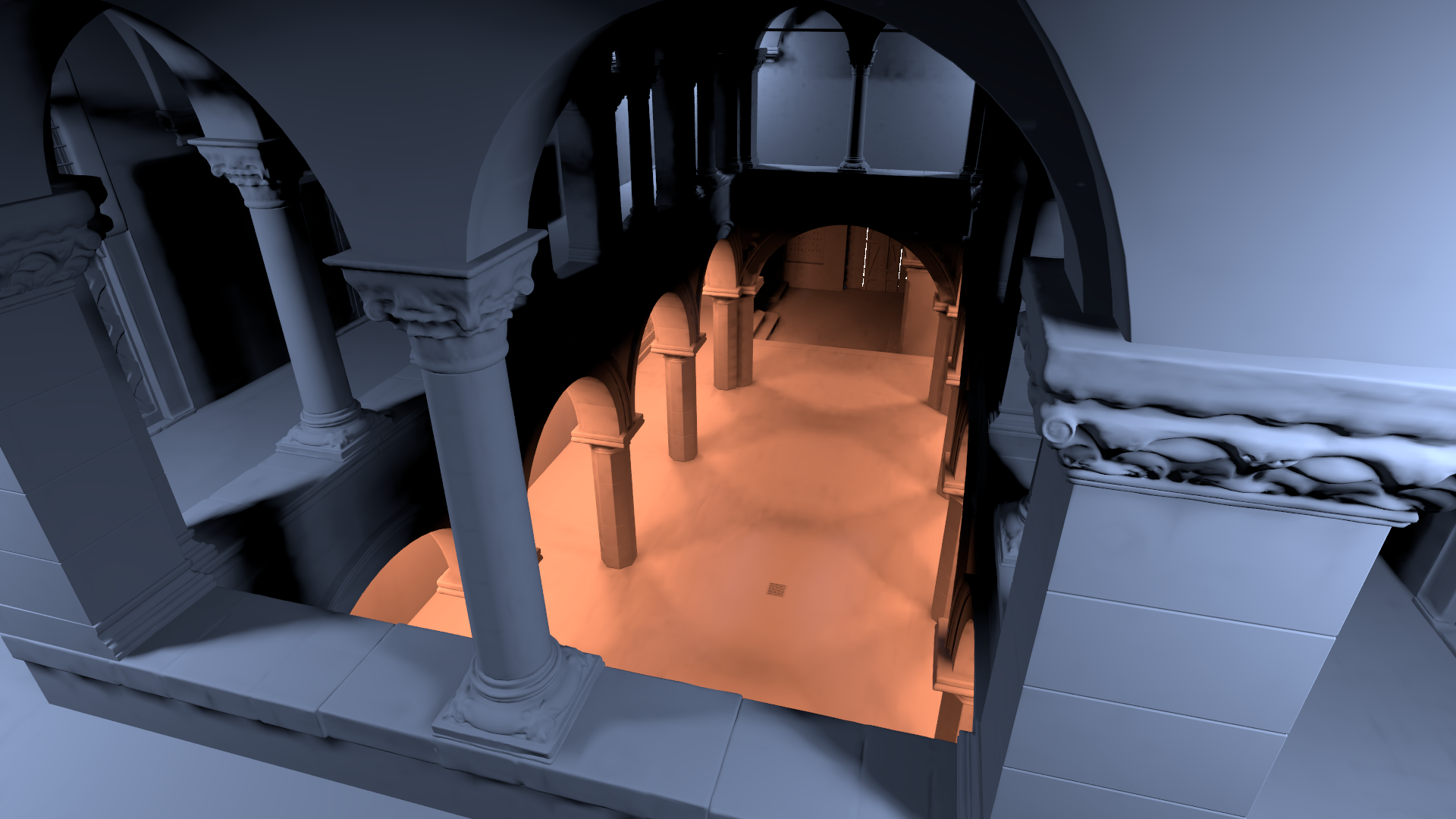}};
        \node[anchor=north west, xshift=2mm, yshift=-2mm, fill=white, opacity=0.7, text opacity=1, font=\scriptsize] 
            at (image.north west) {Neural DI (ours)};
    \end{tikzpicture}
\end{minipage}%
\begin{minipage}{0.485\textwidth}
    \begin{tikzpicture}
        \node[anchor=south west, inner sep=0] (image) at (0,0) {\includegraphics[width=\linewidth]{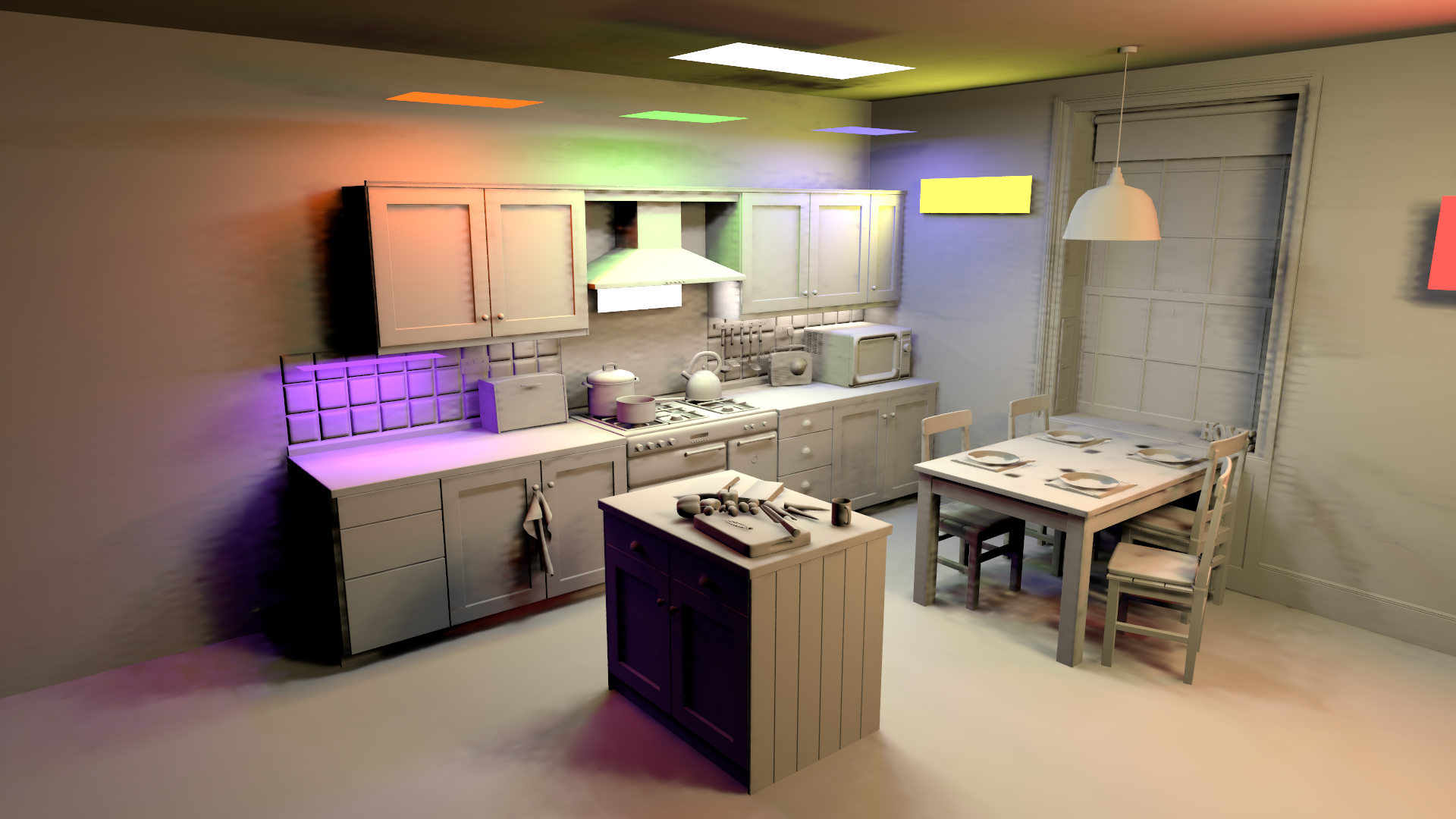}};
        \node[anchor=north west, xshift=2mm, yshift=-2mm, fill=white, opacity=0.7, text opacity=1, font=\scriptsize] 
            at (image.north west) {Neural DI (ours)};
    \end{tikzpicture}
\end{minipage}%

\begin{minipage}{0.485\textwidth}
    \begin{tikzpicture}
        \node[anchor=south west, inner sep=0] (image) at (0,0) {\includegraphics[width=\linewidth]{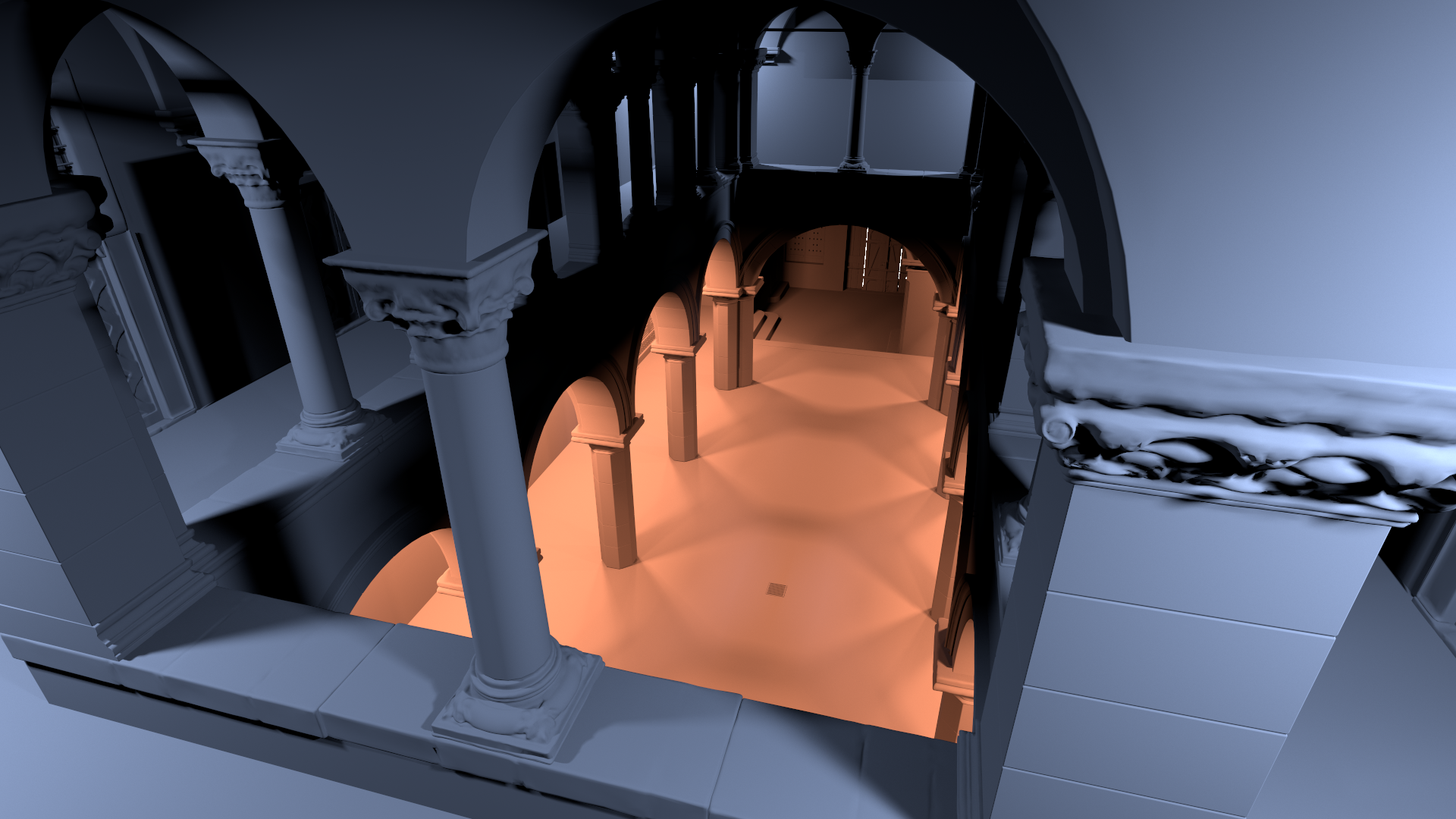}};
        \node[anchor=north west, xshift=2mm, yshift=-2mm, fill=white, opacity=0.7, text opacity=1, font=\scriptsize] 
            at (image.north west) {Ground Truth};
    \end{tikzpicture}
\end{minipage}%
\begin{minipage}{0.485\textwidth}
    \begin{tikzpicture}
        \node[anchor=south west, inner sep=0] (image) at (0,0) {\includegraphics[width=\linewidth]{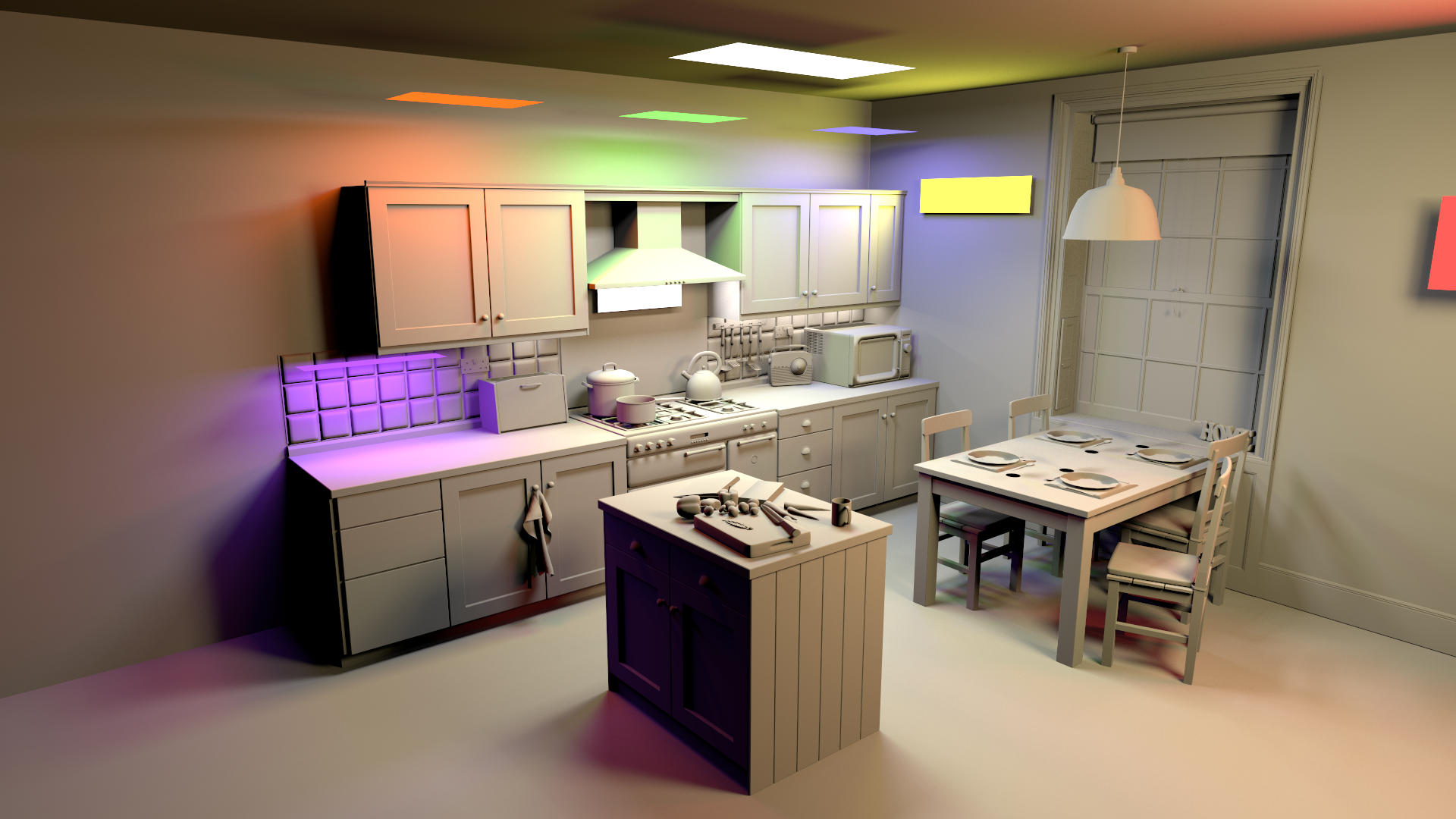}};
        \node[anchor=north west, xshift=2mm, yshift=-2mm, fill=white, opacity=0.7, text opacity=1, font=\scriptsize] 
            at (image.north west) {Ground Truth};
    \end{tikzpicture}
\end{minipage}%

\begin{minipage}{0.485\textwidth}
    \begin{tikzpicture}
        \node[anchor=south west, inner sep=0] (image) at (0,0) {\includegraphics[width=\linewidth]{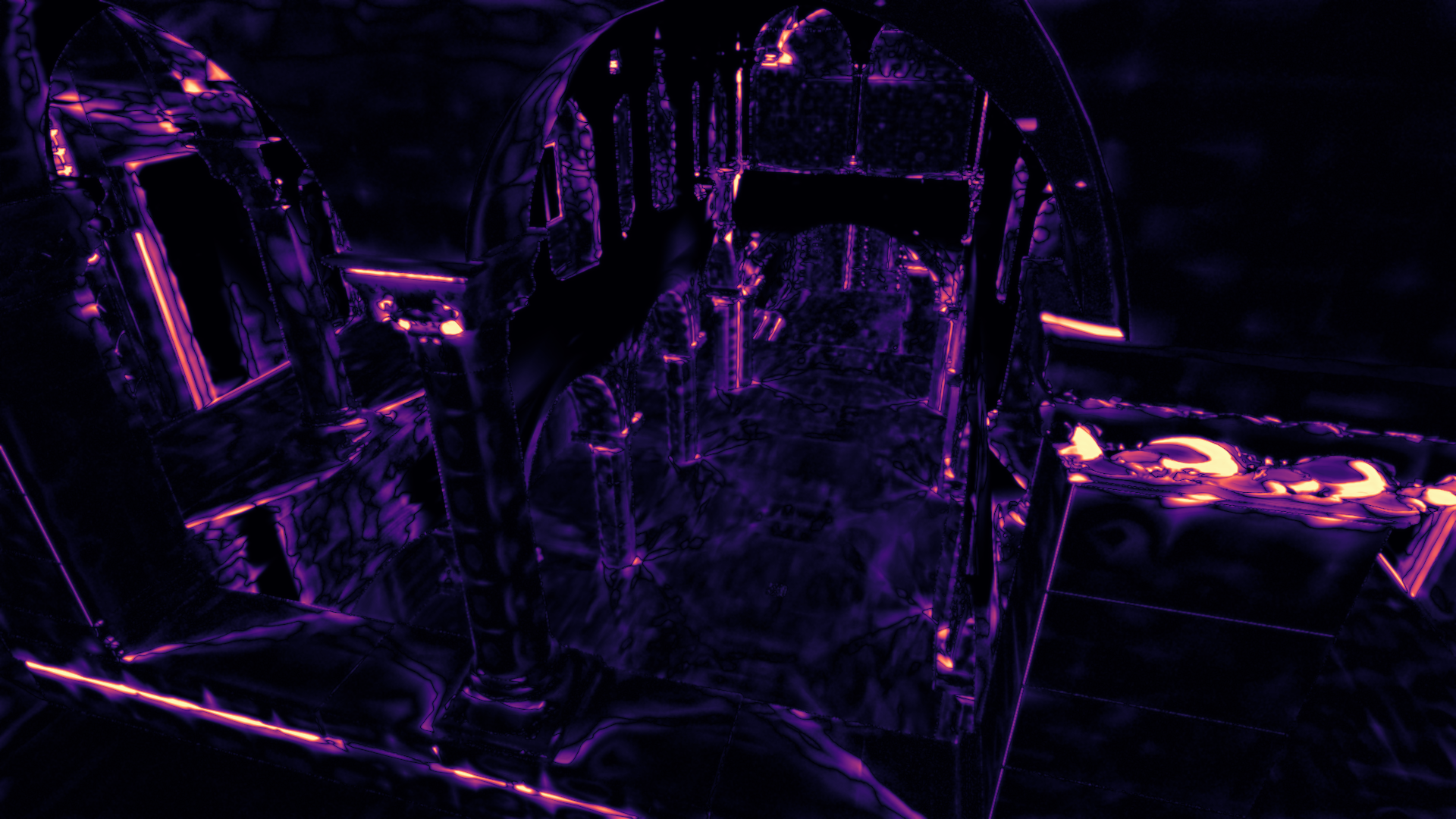}};
        \node[anchor=south east, xshift=-2mm, yshift=2mm, fill=white, opacity=0.7, text opacity=1, font=\scriptsize] 
            at (image.south east) {FLIP 0.067};
        \node[anchor=north west, xshift=2mm, yshift=-2mm, fill=white, opacity=0.7, text opacity=1, font=\scriptsize] 
            at (image.north west) {Neural DI (ours)};
    \end{tikzpicture}
\end{minipage}%
\begin{minipage}{0.485\textwidth}
    \begin{tikzpicture}
        \node[anchor=south west, inner sep=0] (image) at (0,0) {\includegraphics[width=\linewidth]{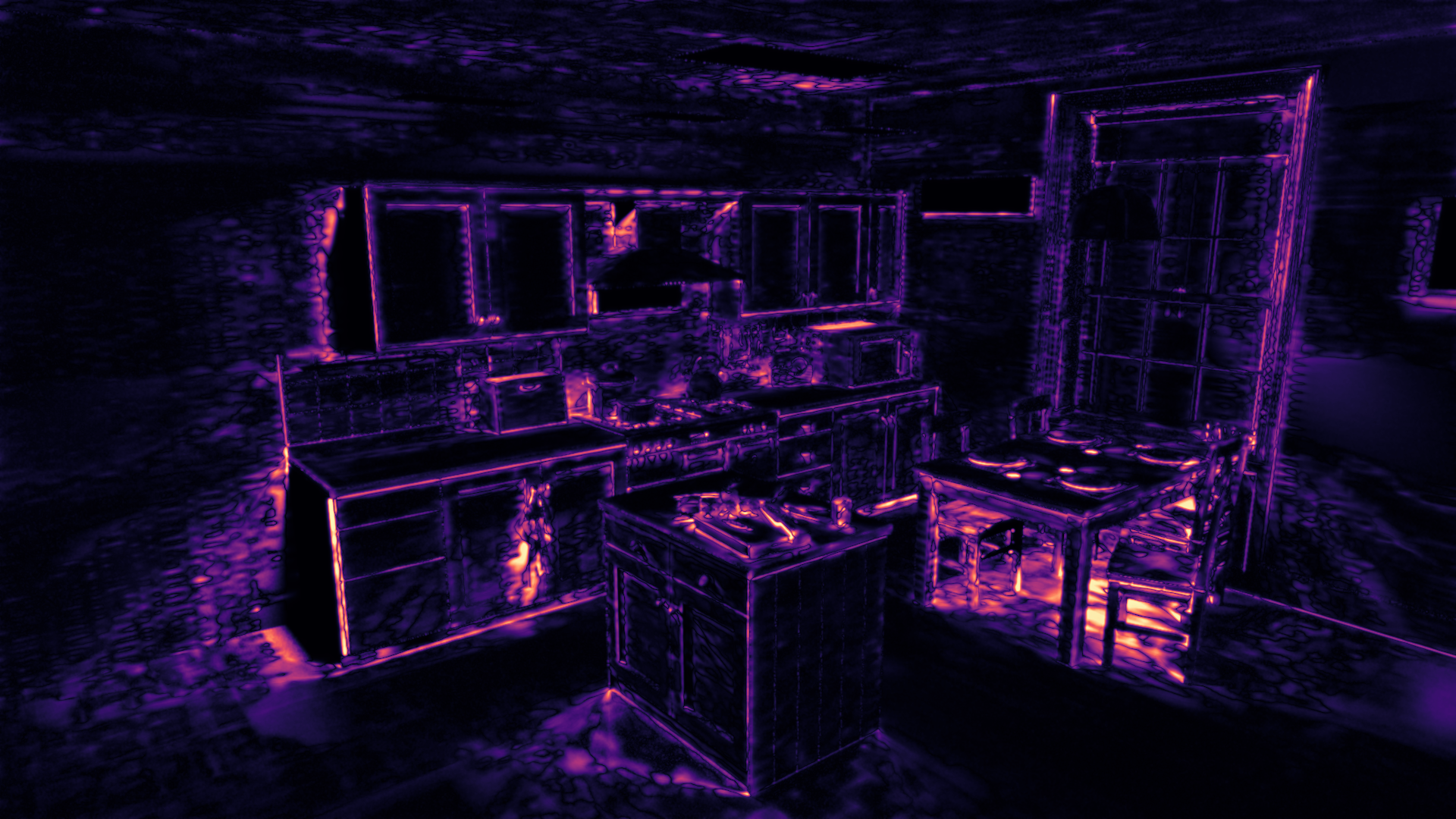}};
        \node[anchor=south east, xshift=-2mm, yshift=2mm, fill=white, opacity=0.7, text opacity=1, font=\scriptsize] 
            at (image.south east) {FLIP 0.083};
        \node[anchor=north west, xshift=2mm, yshift=-2mm, fill=white, opacity=0.7, text opacity=1, font=\scriptsize] 
            at (image.north west) {Neural DI (ours)};
    \end{tikzpicture}
\end{minipage}%

\caption{\label{fig:neural_di}
 Neural DI produces noise-free images using only one sample per pixel: Sponza (left) and kitchen (right) scenes using 32 lights and one sample per pixel. To accentuate shadows, we replaced textured materials with gray diffuse material.  Notice that our method can learn penumbras from noisy training data.
}
\end{figure}

\subsection{Neural Direct Illumination}
\label{sec:NeuralDI}

Since the light weights used for WRS are based on LTC shading and visibility, we can also use them directly as an approximate direct illumination (see Figure~\ref{fig:neural_di}). This yields illumination with approximate shadows, which is biased but very fast and noise-free, without casting any shadow rays. We call this \emph{neural direct illumination} (Neural DI), which can be used as an approximation of direct illumination for deeper bounces in path tracing or for a fast preview. Note that this is only applicable to the case when output neurons represent individual lights, not clusters.

\subsection{Neural Network Architecture}
 
We use a multilayer perceptron (MLP) with two hidden layers, each containing 32 neurons. The activation function for hidden layers is the leaky rectified linear unit (ReLU) with a slope of 0.01. We have tested several activation functions and found that leaky ReLU achieves the lowest training loss. For the output layer, we use the sigmoid activation function. Sigmoid not only maps the output to the range of valid visibility values $[0,1]$, but it also reduces the training loss faster than leaky ReLU. The neural network training with backpropagation uses the L2 loss function. For the multi-resolution hash-grid encoding, we use ten levels with the base resolution 16 and four features per level. This setup leads to approximately 562K learned parameters represented using 32-bit floats. Notice that the majority of these parameters correspond to the hash-grid weights, with only a marginal number dedicated to the MLP. The hash encoding is a critical component in achieving high quality and fast training with our method. For scenes with a low light count, the number of features per level can be decreased.
 
\subsection{Training}

We use the He initialization strategy~\cite{he2015delving} to initialize the MLP. For training, we use the Adam optimizer~\cite{diederik2014adam} with a variable base learning rate. We start with a learning rate of 0.05 and we linearly lower it for the first 200 training steps down to 0.001. This significantly speeds up training at the beginning, which converges to a stable state faster. We perform one training step (one epoch with one batch) per frame. Our training examples consist of a random position in the scene as the input to the network and the corresponding visibility of each light as the target for the network output. 

We have also tried training the network on the shadowed radiance of lights (light intensity attenuated by the squared distance to the light multiplied by visibility). For complex scenes, most of the training samples were close to zero due to strong distance attenuation, and the network had a tendency to predict extremely low values everywhere. Therefore, we settled on training the network only on visibility values. Each training sample comprises mutual visibility for one random sample on a light and a given point. The visibility in this case is binary, but for area lights, the neural network will eventually learn the average visibility over the whole area of the light (penumbra), which is not binary.

\begin{figure}[b]
\centering
\begin{minipage}{0.485\textwidth}
    \begin{tikzpicture}
        \node[anchor=south west, inner sep=0] (image) at (0,0) {\includegraphics[width=\linewidth]{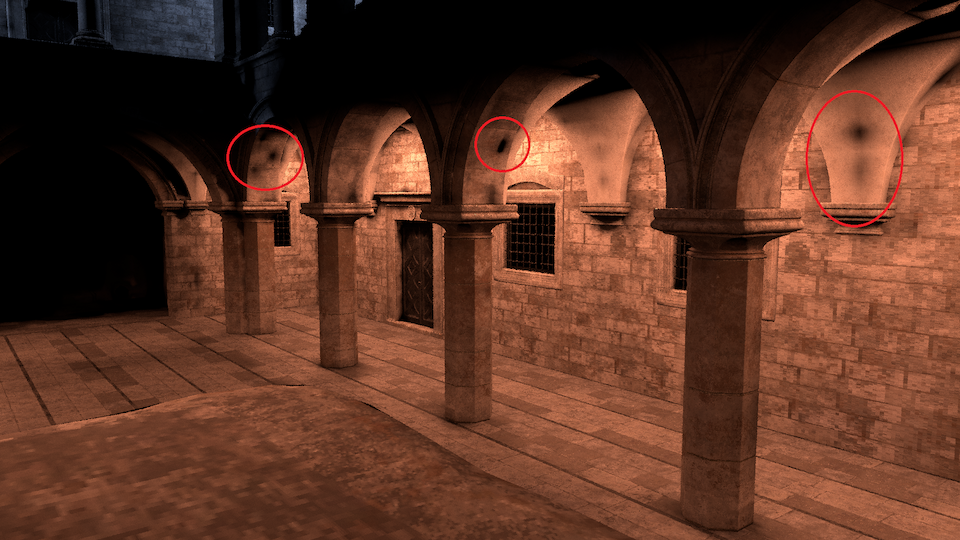}};
        \node[anchor=south east, xshift=-2mm, yshift=2mm, fill=white, opacity=0.7, text opacity=1, font=\scriptsize] 
            at (image.south east) {World-space data};
    \end{tikzpicture}
\end{minipage}%
\begin{minipage}{0.485\textwidth}
    \begin{tikzpicture}
        \node[anchor=south west, inner sep=0] (image) at (0,0) {\includegraphics[width=\linewidth]{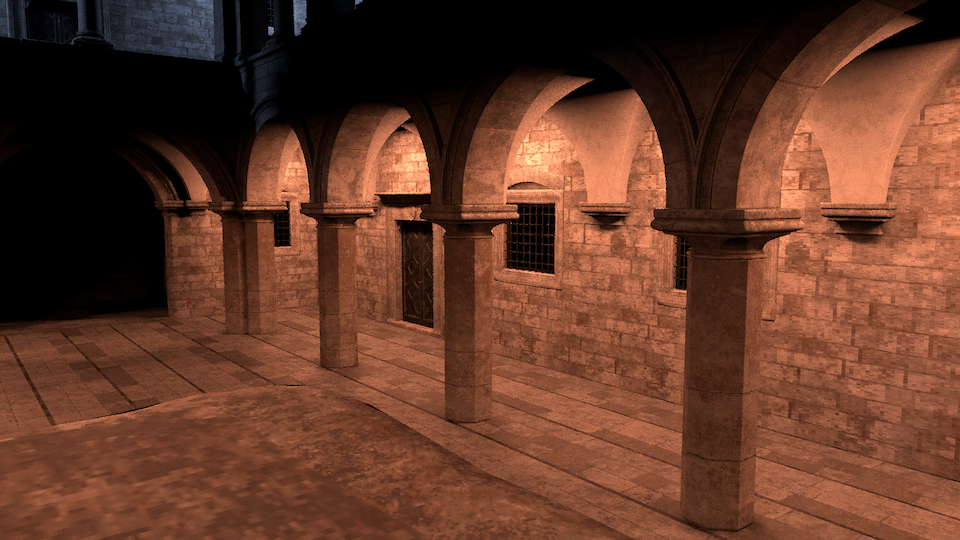}};
        \node[anchor=south east, xshift=-2mm, yshift=2mm, fill=white, opacity=0.7, text opacity=1, font=\scriptsize] 
            at (image.south east) {World-space + screen-space data};
    \end{tikzpicture}
\end{minipage}%
\caption{\label{fig:dark_blobs}
 Using world-space data for training causes artifacts that manifest as dark blobs, highlighted in red circles (left). Introducing screen-space data for training fixes the problem (right). Images show our biased Neural DI method.}
\end{figure}

There are several options for generating these training examples. Random points on surfaces visible from the camera (screen space) achieve best result for a given camera view, but the network adapts slowly for new views. Using random points within scene bounds (world space) needs no adaptation for new views, but introduces dark blob artifacts (see Figure~\ref{fig:dark_blobs}). A third option, generating samples on the geometry, does not train the network to predict visibility in empty areas, which might become occupied in future frames due to animation. Additionally, it makes it impossible to use our method for light sampling within volumes of participating media, so we do not recommend it. We found that using a combination of world-space data and screen-space data works best as it fixes the artifacts but also adapts rapidly to camera movement. Our solution uses a combination of 4096 world-space samples and another 4096 screen-space samples.

We cast a shadow ray toward each of the light sources from each training point to produce the training example. This is very fast in practice, since we only use 8196 training points per frame, and the number of lights is limited to the number of output neurons (32 in our tests).

\subsection{Dynamic Scenes}
\label{sec:dynamic_scenes}

Our method is online-trained, therefore it supports dynamic scenes including animated geometry, lights, and camera. When the scene changes, the network state might not approximate the visibility well until it adapts to the new situation. As our method is unbiased, this exhibits as an increase in variance (noise) but not bias. When training from scratch (see Figure~\ref{fig:training_steps}), the training needs only about 16 frames to reduce the FLIP error to almost a half. Smooth changes to the scene can be expected to adapt faster than 16 frames, but the application may need to increase the learning rate temporarily, or perform more than one training step per frame. Abrupt changes, such as teleportation of camera, or dynamic geometry are handled by using world-space samples that pre-train the network to be used on unseen views and previously unoccupied space.

\subsection{Clustering}
\label{sec:clustering}

The method described so far, representing visibility of each light in the scene with a dedicated output neuron, limits the number of supported lights to the size of the output layer allowed by the selected network architecture and target hardware (32 in our implementation). In this section, we introduce a \emph{clustered neural visibility cache} approach (Clustered NVC), where instead of representing individual lights, the output neurons represent the visibility of a cluster, which can consist of an arbitrary number of lights.

We use the $k$-means algorithm~\cite{macqueen1967some, lloyd1982least} to cluster the lights into $k$ clusters (we use $k=32$). During training, we randomly select a light source within each cluster to train the network to predict average visibility of the cluster for any point in the scene. This approach works especially well for interior scenes with many rooms, as we can quickly cull light clusters not contributing to the room with the camera. Alternatively, lights can be clustered based on the mesh they belong to (e.g., clustering emissive triangles of a complex mesh representing a lamp or a neon~sign).

To sample a light using our neural visibility cache with clusters, we employ a two-step process based on WRS and resampled importance sampling (RIS)~\cite{talbot2005importance} implemented with reservoirs~\cite{bitterli20spatiotemporal}. The first step uses WRS to sample a cluster $y$ out of $m$ clusters, based on the inferred average cluster visibility at that point, resulting in a reservoir with the selected sample $y$, its sampling weight $\mathrm{w}(y)$, and the sum of their sampling weights $\mathrm{w}_{\text{sum}} = \sum_{j=1}^m \mathrm{w}(y_j)$. The second step uses a streaming RIS~\cite{bitterli20spatiotemporal} to generate a final light sample $x$ selected from the pool of $m_y$ lights within the cluster $y$. The source probability density function $p(x)$ for RIS is defined as $p(x)= m\frac{\mathrm{w}(y)}{\mathrm{w}_{\text{sum}}}\frac{1}{m_y}$, where $m\frac{\mathrm{w}(y)}{\mathrm{w}_{\text{sum}}}$ is a reciprocal weight of the reservoir from the first step and $\frac{1}{m_y}$ is the probability of sampling a light uniformly within the cluster. The target probability density function $\hat{p}(x)$ for RIS is selected based on the BRDF contribution and LTC lighting, same as before.

We found out that more training data is needed to provide plausible results for scenes with many lights. Therefore, we use 49,152 training examples for the Subway scene with 59K lights. We have also found out that the optimal hash-grid settings are different for the clustered approach, since the approximation we are trying to learn is lower frequency in nature, and we use eight levels with the base resolution 2 for this~case.

\section{Results and Discussion}

We implemented the proposed method in DirectX 12 and HLSL. Rendering each frame consists of the following five passes: (1)~G-buffer generation, (2)~training data generation, (3)~network training, (4)~inference and light sampling, and (5)~shading. The neural network is implemented as a fully fused MLP~\cite{muller2021real}, allowing us to train and infer the neural network entirely in the on-chip shared memory. Our implementation allows to execute inference inline in the scope of DirectX Raytracing (DXR) ray tracing shaders, enabling to use our method on every bounce of light while tracing paths. All tests were executed on AMD Radeon RX 7900 XT.

\begin{figure*}[!h]
\begin{center}
\scalebox{1.0}{
\scriptsize
\begin{minipage}{\textwidth}
\centering

\begin{minipage}[t]{1.0\textwidth}
\centering
\setlength{\tabcolsep}{1pt}
\renewcommand{\arraystretch}{0.5}
\begin{tabular}{ccccccc}
& 1 & 8 & 16 & 32 & 64 & 128 \\

\rotatebox{90}{NRC} &
\includeimgcrop{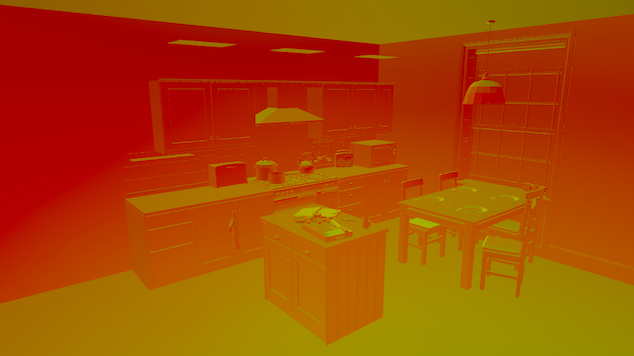} &
\includeimgcrop{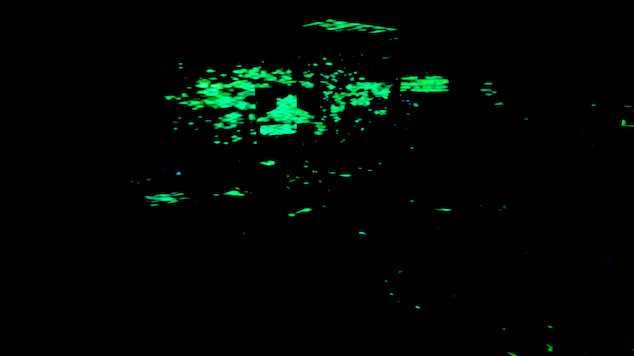} &
\includeimgcrop{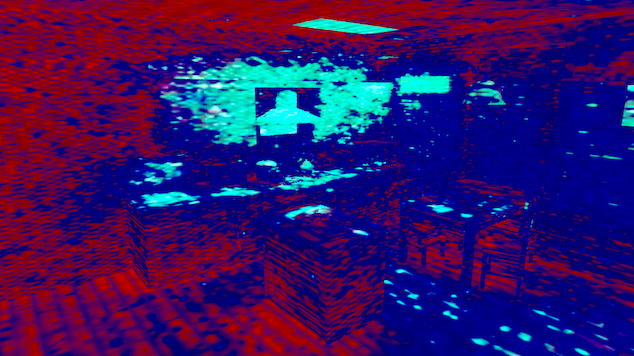} &
\includeimgcrop{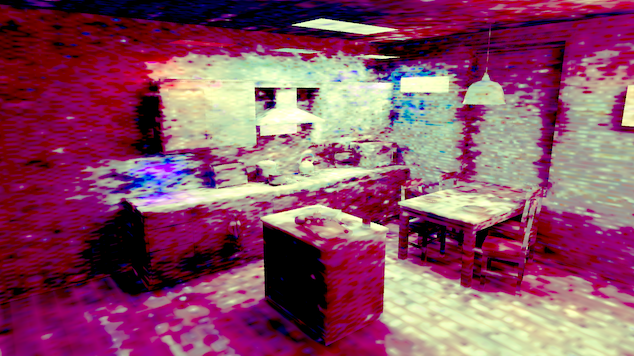} &
\includeimgcrop{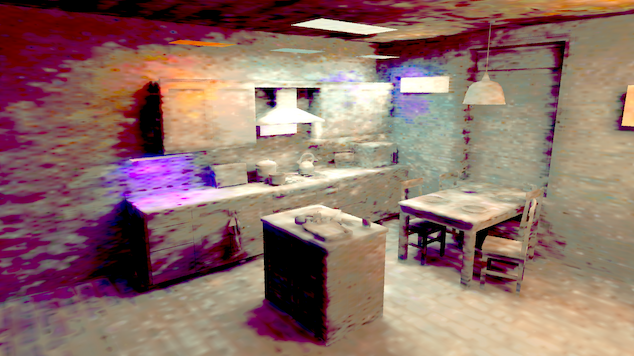} &
\includeimgcrop{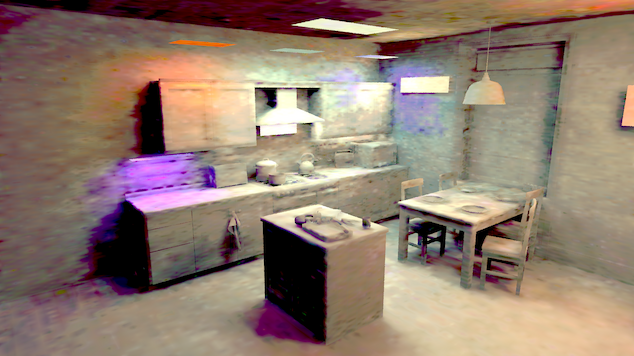} \\
& 0.808 & 0.952 & 0.980 & 0.724 & 0.450 & 0.351 \\
\rotatebox{90}{NLS (ours)} &
\includeimgcrop{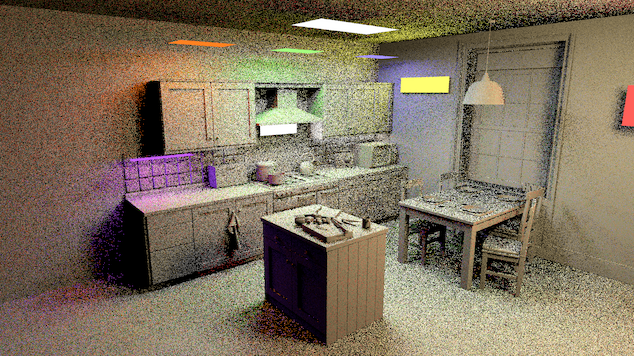} &
\includeimgcrop{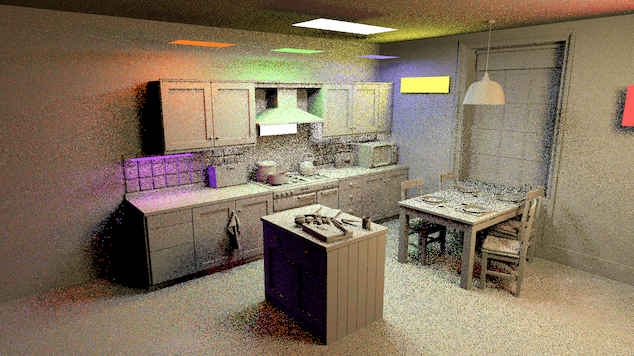} &
\includeimgcrop{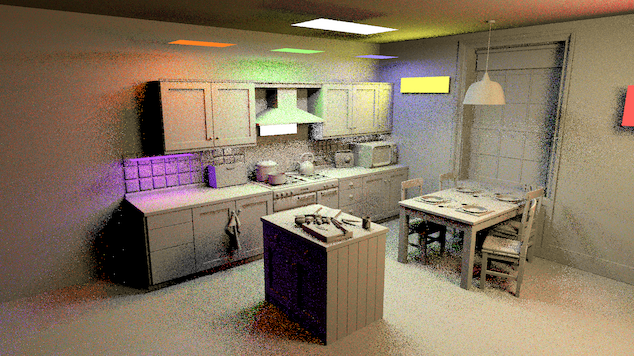} &
\includeimgcrop{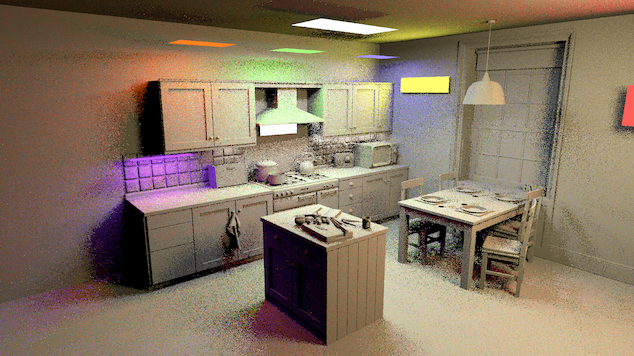} &
\includeimgcrop{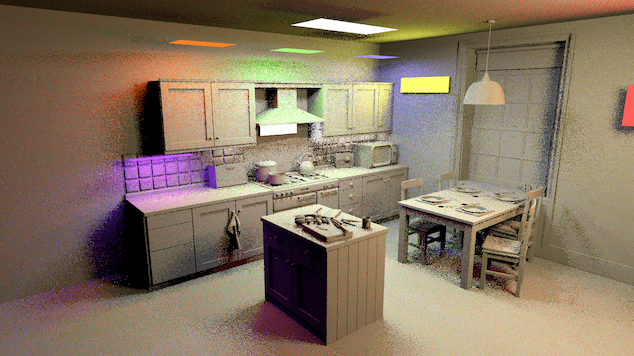} &
\includeimgcrop{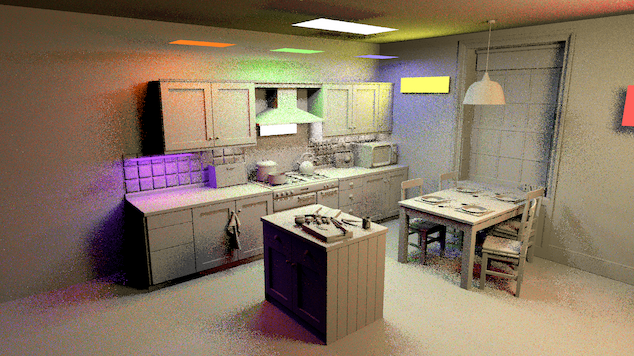} \\
& 0.566 & 0.385 & 0.348 & 0.318 & 0.305 & 0.298 \\
\rotatebox{90}{ReSTIR} &
\includeimgcrop{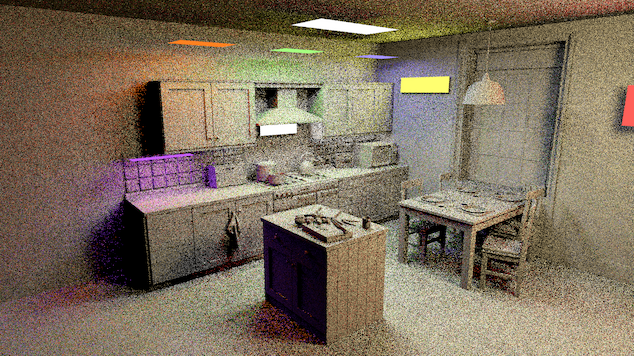} &
\includeimgcrop{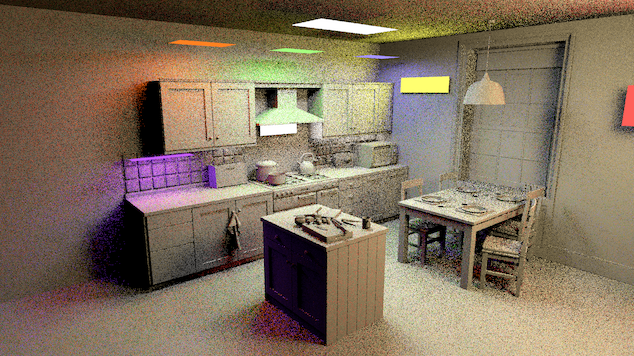} &
\includeimgcrop{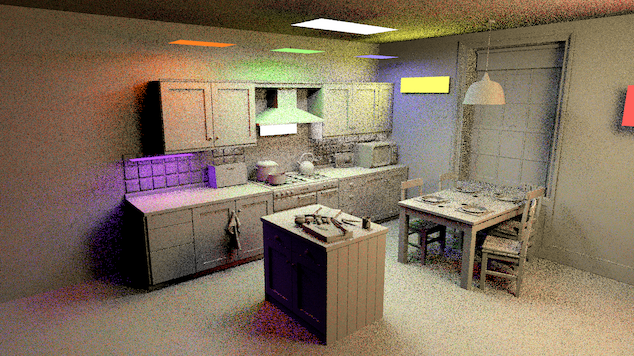} &
\includeimgcrop{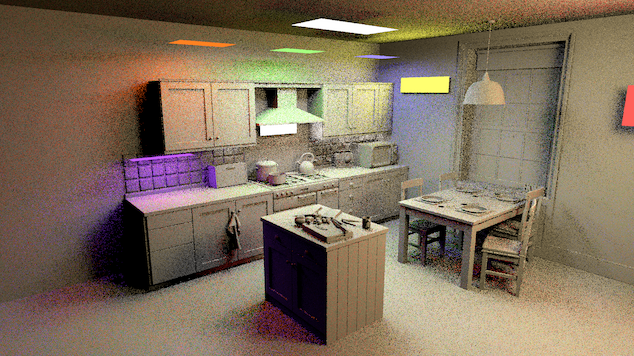} &
\includeimgcrop{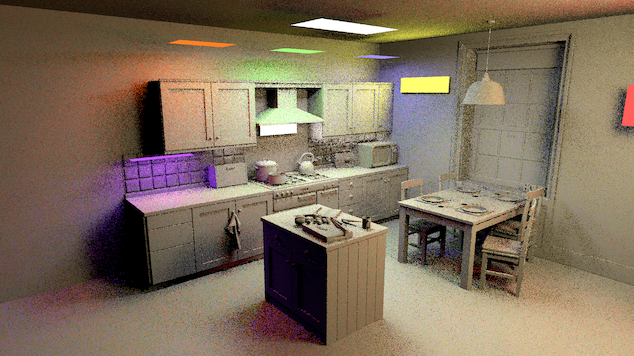} &
\includeimgcrop{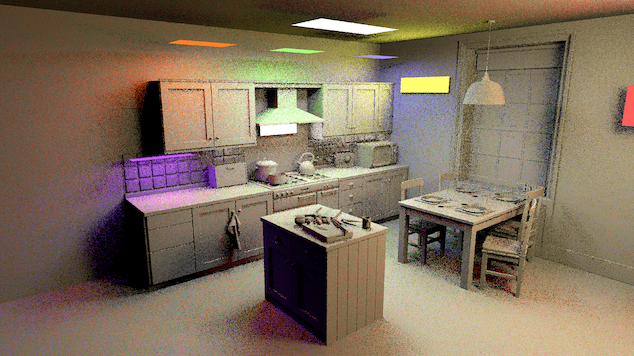} \\
& 0.575 & 0.444 & 0.423 & 0.393 & 0.369 & 0.361 \\
\end{tabular}
\end{minipage}
\begin{minipage}[t]{1.0\textwidth}
\centering
\begin{tabular}{cc}
\includegraphics[height=0.158\textheight]{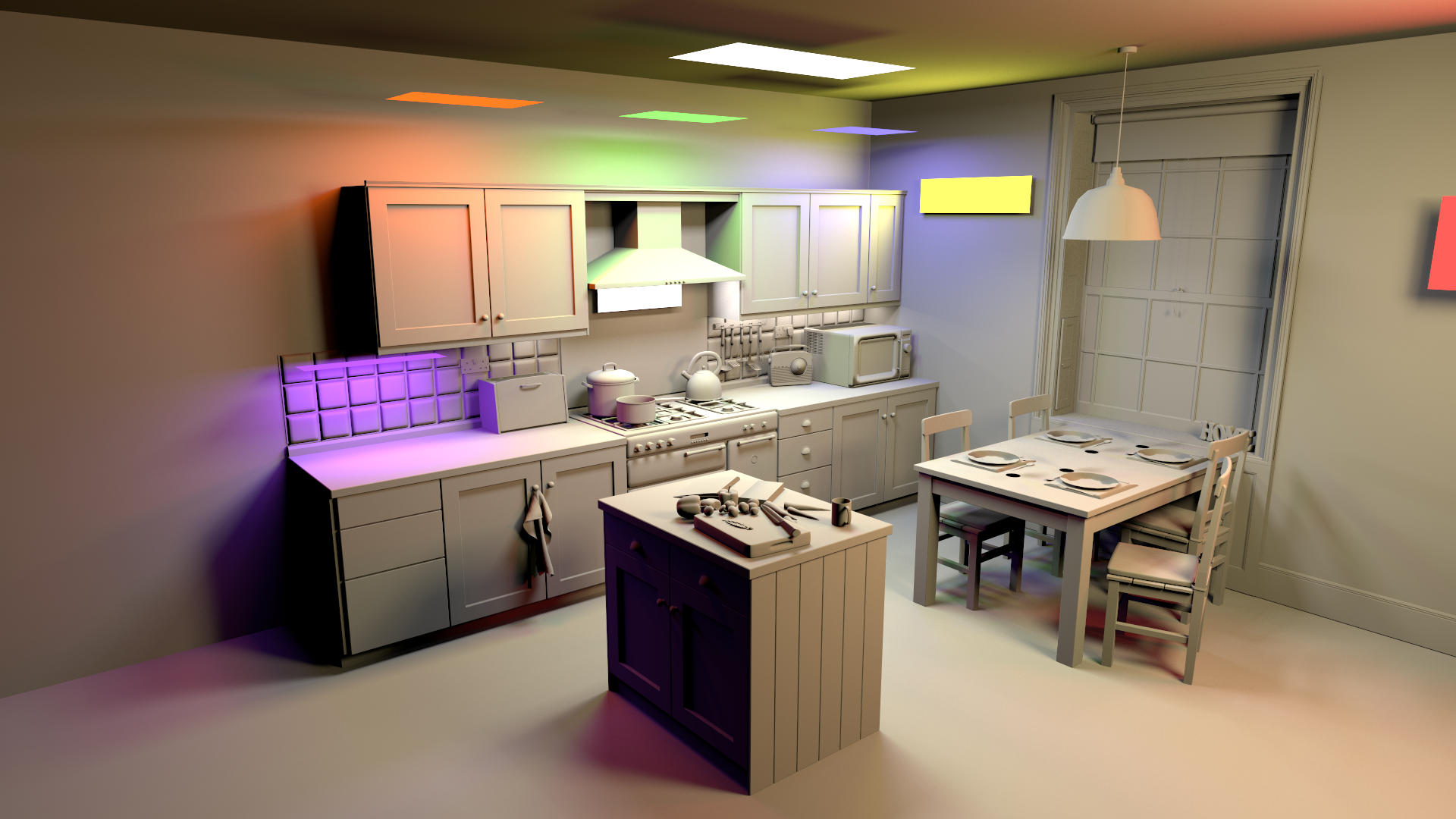} &
\includegraphics[height=0.158\textheight]{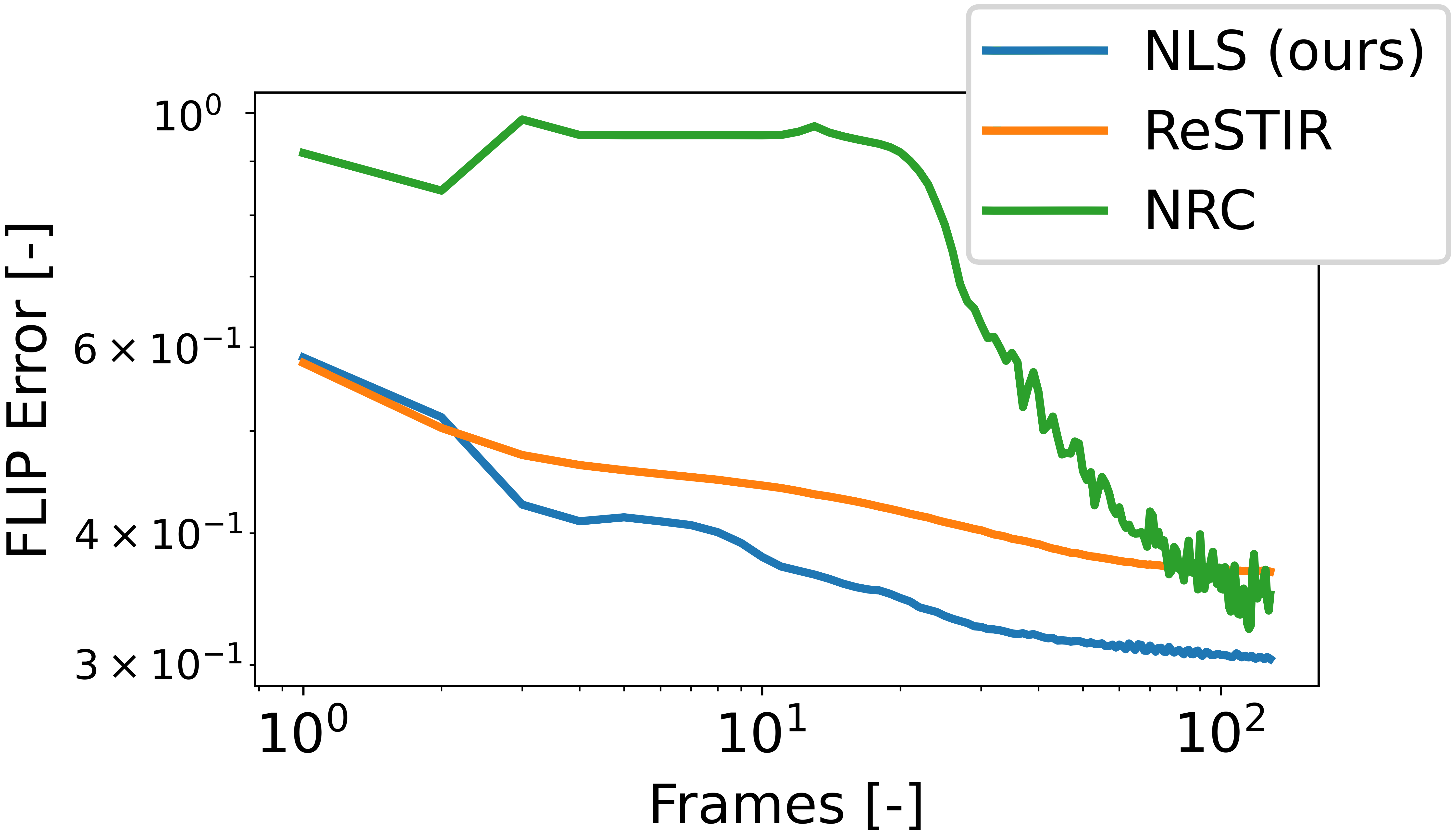} \\
Ground Truth & FLIP Over Time
\end{tabular}
\end{minipage}

\end{minipage}
}

\end{center}
\caption{A comparison of NRC (top), our neural light sampling (NLS) (middle) and screen space ReSTIR (bottom) on a progression of 128 frames. The number under each image represents its FLIP error, which decreases over time for all methods, but NRC suffers from color artifacts, especially at the beginning of training. Our method remains unbiased even at the beginning while the network is not sufficiently trained, exhibiting as increased variance, similarly to ReSTIR. Our method reduces the FLIP error faster than other methods, achieving lower error than ReSTIR even only after three training steps.}
\label{fig:training_steps}
\end{figure*}

\subsection{Comparison to Screen-Space ReSTIR}
\label{sec:Comparison_Restir}

As a reference method, we implemented screen-space ReSTIR using eight initial candidates, casting a shadow ray for the selected initial candidate. We use temporal resampling where we clamp the contribution of the previous reservoir to $20\times$ the contribution of the new reservoir. Spatial resampling uses a radius of 32 pixels. Unless stated otherwise, we use one sample per pixel for all tests, and both ReSTIR and the neural network are converged for 1024 frames. Note that the FLIP error of both methods is already significantly reduced after 20 to 30 frames (see Figure~\ref{fig:training_steps}).

\begin{figure}
\centering
\begin{tikzpicture}
    \node[anchor=south west, inner sep=0] (image) at (0,0) {
        \includegraphics[width=0.9\linewidth, trim=0 2.5cm 0 2.5cm, clip]{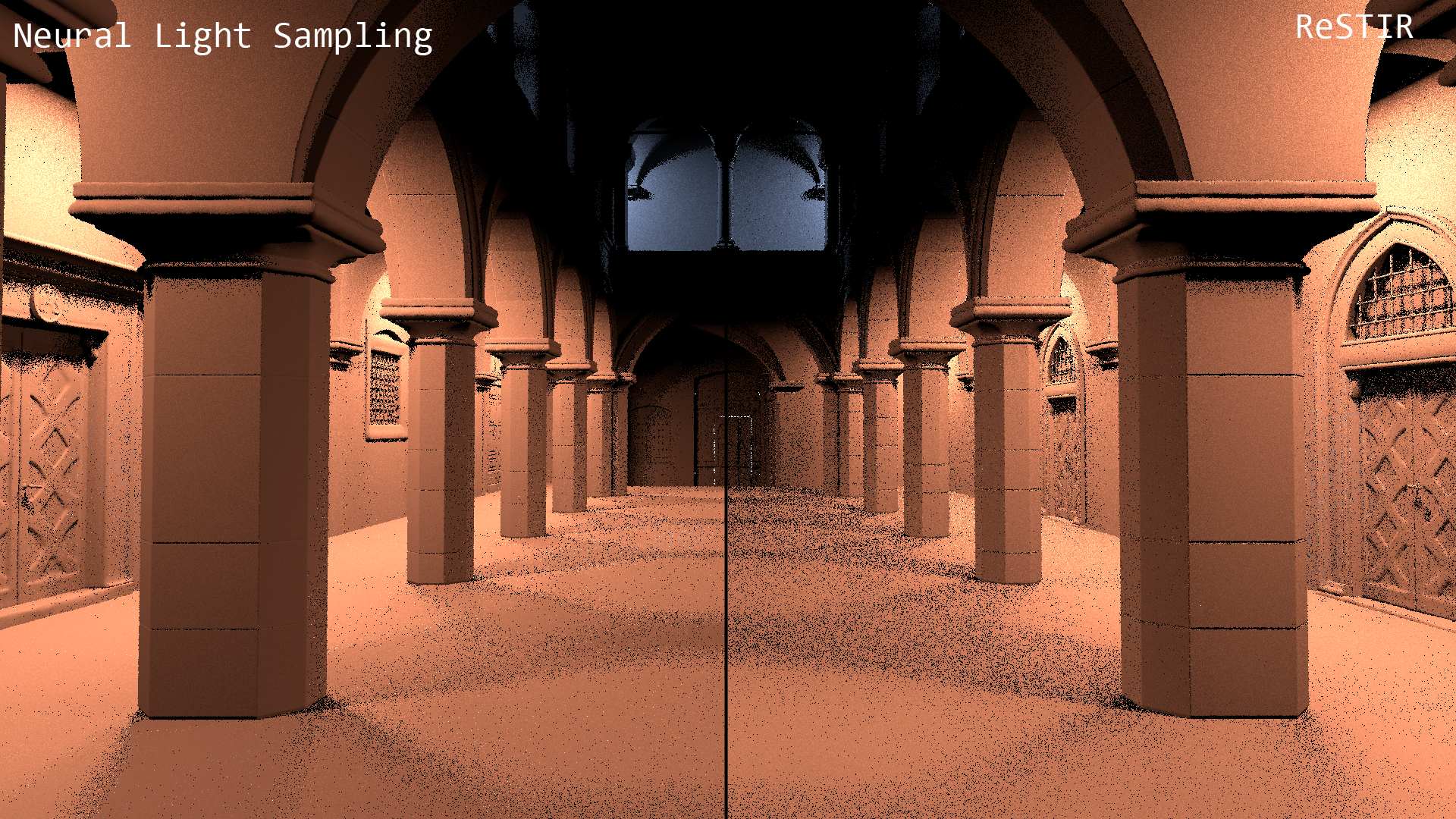}
    };
    \node[anchor=south east, xshift=-2mm, yshift=2mm, fill=white, opacity=0.7, text opacity=1, font=\scriptsize] 
        at (image.south east) {FLIP 0.195};
    \node[anchor=north east, xshift=-2mm, yshift=-2mm, fill=white, opacity=0.7, text opacity=1, font=\scriptsize] 
        at (image.north east) {ReSTIR};
    \node[anchor=north west, xshift=2mm, yshift=-2mm, fill=white, opacity=0.7, text opacity=1, font=\scriptsize] 
        at (image.north west) {Neural Light Sampling (ours)};
    \node[anchor=south west, xshift=2mm, yshift=2mm, fill=white, opacity=0.7, text opacity=1, font=\scriptsize] 
        at (image.south west) {FLIP 0.103};
\end{tikzpicture}
\caption{Comparison of neural light sampling (left) to screen-space ReSTIR (right) on the Sponza scene with 32 lights. Neural light sampling produces lower error at the same performance cost.}
\label{fig:restir_vs_nls}
\end{figure}

Compared to screen-space ReSTIR, our neural light sampling achieves a lower FLIP error~\cite{Andersson2021b} at a similar time budget, especially in the occluded regions (see Figures~\ref{fig:teaser} and \ref{fig:restir_vs_nls}). FLIP error is reduced by about $20\%$ for the Kitchen scene and about $45\%$ for the Sponza scene. We can make the method unbiased by clamping the output of the neural network (discussed in Section~\ref{sec:Outline}) at the cost of slightly increased variance (see Figure~\ref{fig:bias}). Neural DI that directly uses the visibility estimates provided by the neural network (see Section~\ref{sec:NeuralDI}) is biased by definition, yet it provides significantly lower error than ReSTIR and neural light sampling (see Figure~\ref{fig:teaser}). The bias exhibits at the edges and in heavily shadowed areas (see Figure~\ref{fig:neural_di}).

\subsection{Combining ReSTIR with Clustered NVC}
\label{sec:restir_with_NVC}

\begin{figure*}
\centering
\begin{minipage}{0.33\textwidth}
    \begin{tikzpicture}
        \node[anchor=south west, inner sep=0] (image) at (0,0) {\includegraphics[width=\linewidth]{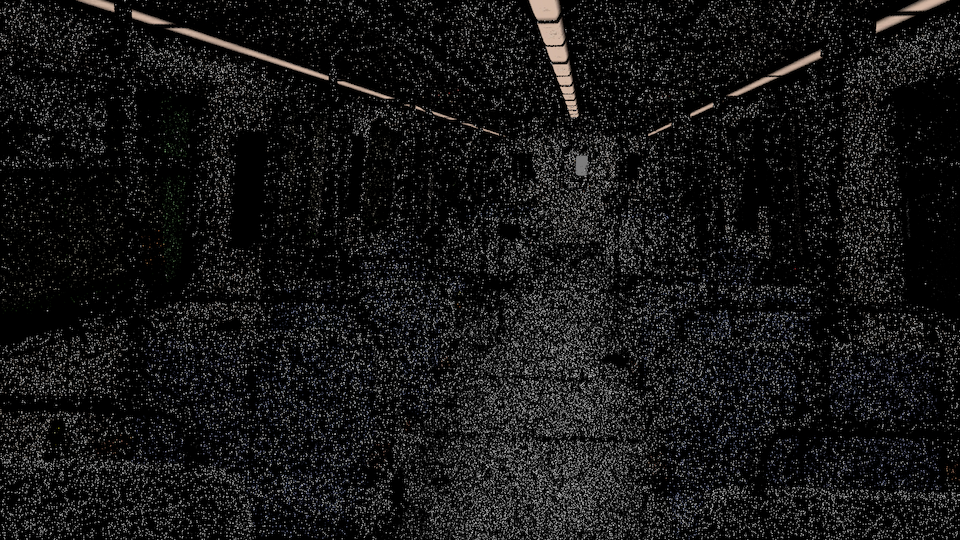}};
        \node[anchor=south east, xshift=-1mm, yshift=1mm, fill=white, opacity=0.7, text opacity=1, font=\tiny] 
            at (image.south east) {FLIP 0.876};
        \node[anchor=north west, xshift=1mm, yshift=-1mm, fill=white, opacity=0.7, text opacity=1, font=\tiny] 
            at (image.north west) {ReSTIR};
    \end{tikzpicture}
\end{minipage}%
\begin{minipage}{0.33\textwidth}
    \begin{tikzpicture}
        \node[anchor=south west, inner sep=0] (image) at (0,0) {\includegraphics[width=\linewidth]{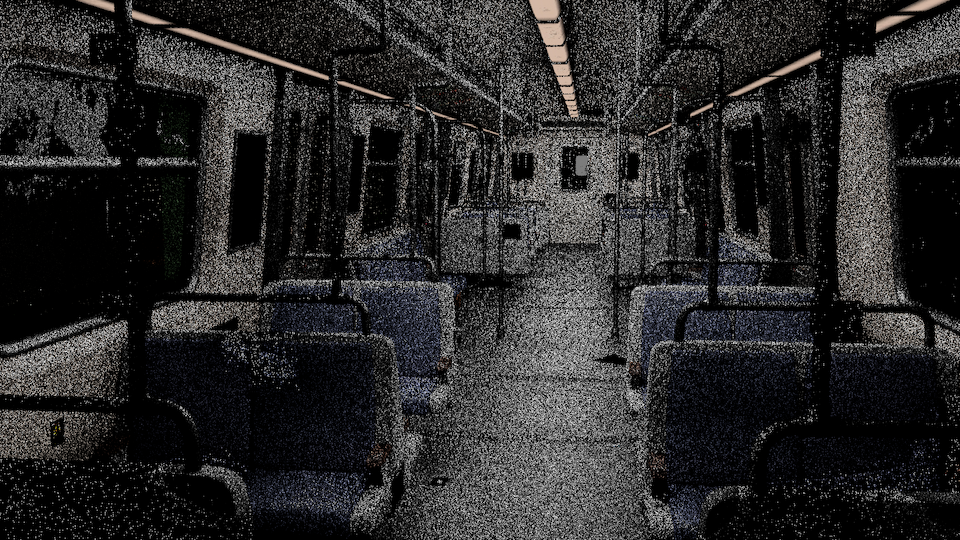}};
        \node[anchor=south east, xshift=-1mm, yshift=1mm, fill=white, opacity=0.7, text opacity=1, font=\tiny] 
            at (image.south east) {FLIP 0.802};
        \node[anchor=north west, xshift=1mm, yshift=-1mm, fill=white, opacity=0.7, text opacity=1, font=\tiny] 
            at (image.north west) {C-NVC$\rightarrow$ReSTIR};
    \end{tikzpicture}
\end{minipage}%
\begin{minipage}{0.33\textwidth}
    \begin{tikzpicture}
        \node[anchor=south west, inner sep=0] (image) at (0,0) {\includegraphics[width=\linewidth]{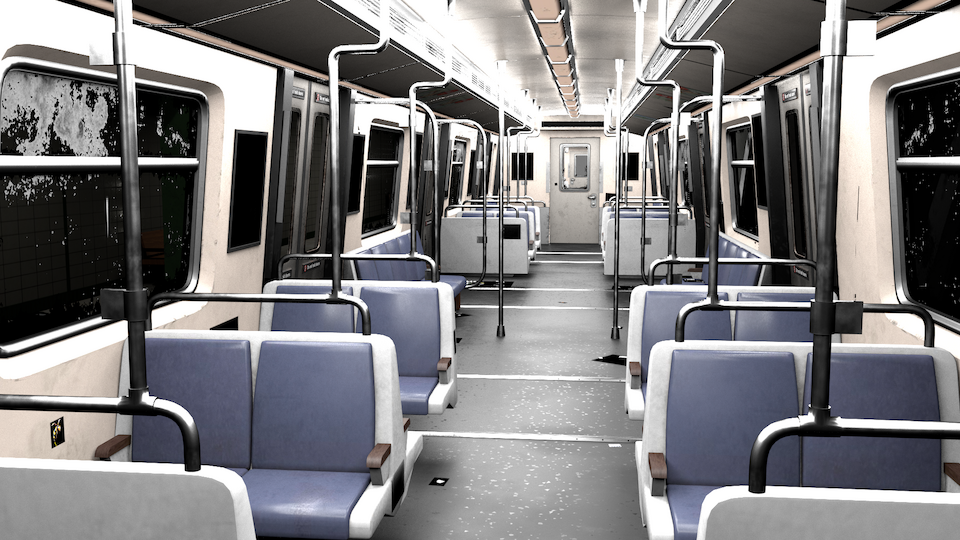}};
        \node[anchor=north west, xshift=1mm, yshift=-1mm, fill=white, opacity=0.7, text opacity=1, font=\tiny] 
            at (image.north west) {Ground Truth};
    \end{tikzpicture}
\end{minipage}%
\caption{\label{fig:cluster_restir}
 A comparison of standard ReSTIR (left), ReSTIR with initial candidates generated by our clustered NVC (center), and the ground truth (right) in an idealized case where disocclusions happen in every pixel. The Subway scene contains 59K lights. We simulate disocclusions by invalidating motion vectors for all pixels, using one sample per pixel. This demonstrates how clustered NVC can help ReSTIR to recover after disocclusion.}
\end{figure*}

To generate initial candidates, standard ReSTIR uses a RIS loop with a user-selected number of candidates (we use eight), which generates a reservoir that can participate in the streaming RIS and for spatial and temporal reuse. We combine our neural approach with ReSTIR by replacing the initial candidates generation routine with our clustered NVC. Our method also generates a reservoir, using a two-step process described in Section~\ref{sec:clustering}, therefore it can be directly used to generate initial candidates.

\begin{figure*}[p]
\centering
\scalebox{1.06}{
\scriptsize
\begin{minipage}{\textwidth}
\centering

\hspace*{-27pt}\begin{minipage}[t]{1.0\textwidth}
\centering
\setlength{\tabcolsep}{1pt}
\renewcommand{\arraystretch}{0.9}
\begin{tabular}{cccc}
& Subway & ZeroDay & Bistro Exterior \\
& 59,164 lights & 4820 lights & 4499 lights \\

\rotatebox{90}{RIS} &
\includeimgcroplarge{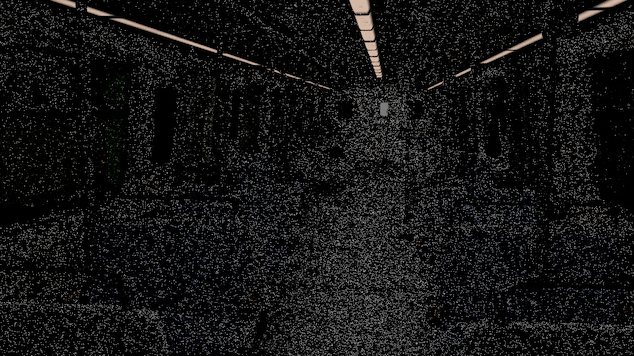} &
\includeimgcroplarge{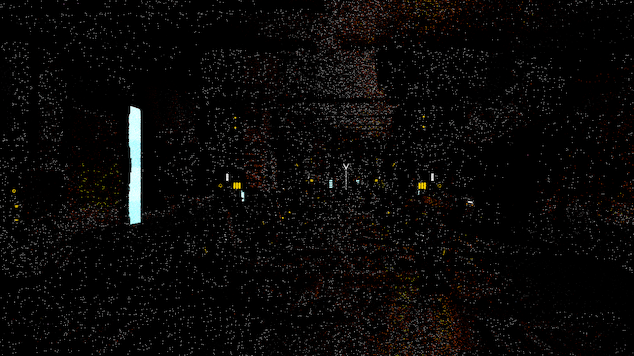} &
\includeimgcroplarge{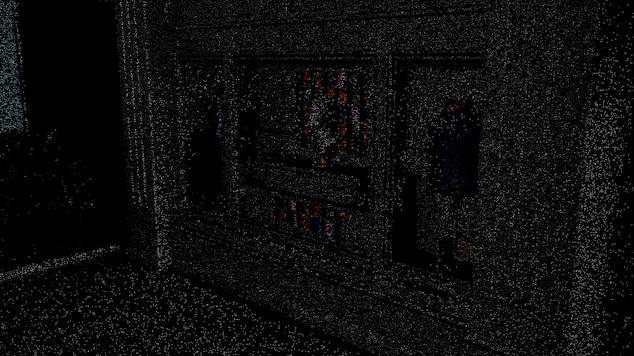} \\
& 0.880 & 0.828 & 0.836 \\
& 2.92~ms & 2.00~ms & 2.74~ms \\
\rotatebox{90}{C-NVC (ours)} &
\includeimgcroplarge{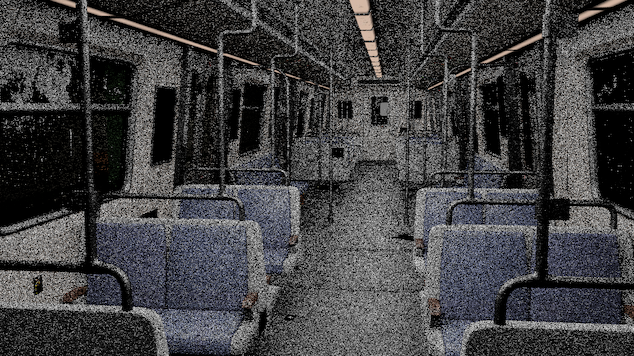} &
\includeimgcroplarge{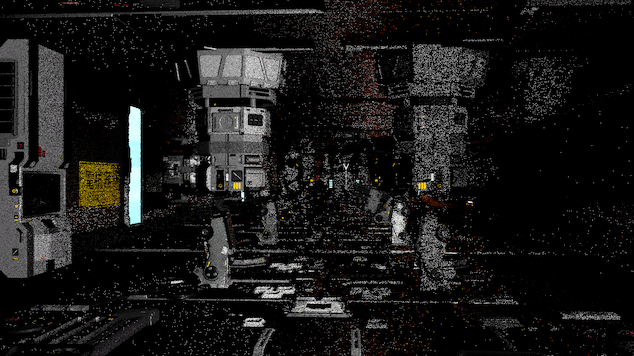} &
\includeimgcroplarge{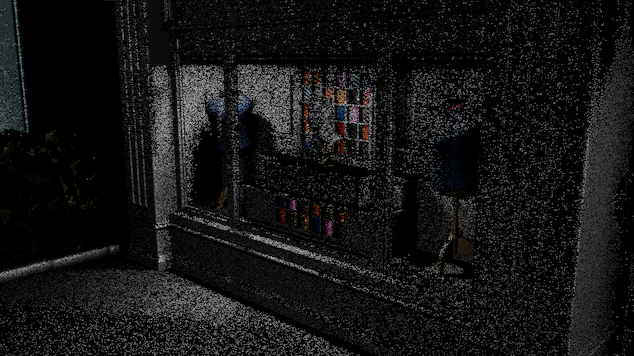} \\
& 0.768 & 0.754 & 0.790 \\
& 7.37~ms & 6.18~ms & 7.85~ms \\
\rotatebox{90}{ReSTIR} &
\includeimgcroplarge{24_subway-NVC-HashGrid-AdaptiveR-F-4-RESTIR_only-frames-1024.pfm.png} &
\includeimgcroplarge{25_zeroday-NVC-HashGrid-AdaptiveR-F-4-RESTIR_only-frames-1024.pfm.png} &
\includeimgcroplarge{26_bistro_exterior_night-NVC-HashGrid-AdaptiveR-F-4-RESTIR_only-frames-1024.pfm.png} \\
& 0.500 & 0.580 & 0.445 \\
& 4.48~ms & 3.43~ms & 5.15~ms \\
\rotatebox{90}{C-NVC$\rightarrow$ReSTIR} &
\includeimgcroplarge{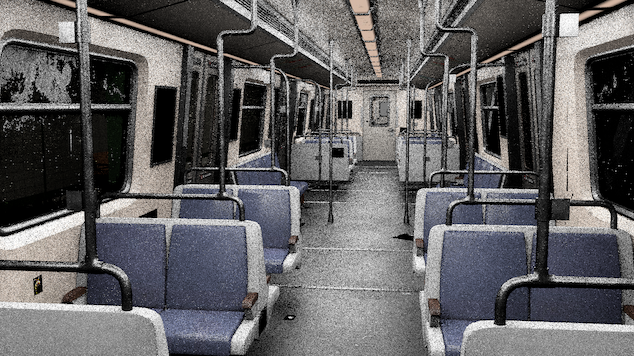} &
\includeimgcroplarge{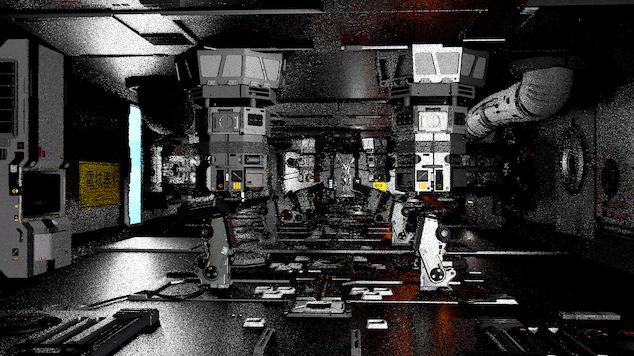} &
\includeimgcroplarge{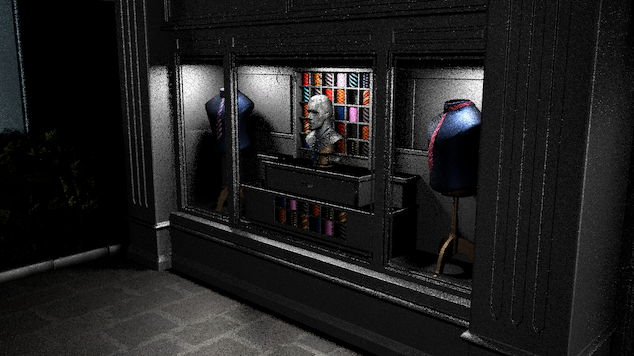} \\
& 0.464 & 0.569 & 0.423 \\
& 8.99~ms & 7.75~ms & 8.24~ms \\
\rotatebox{90}{Ground Truth} &
\includeimgcroplarge{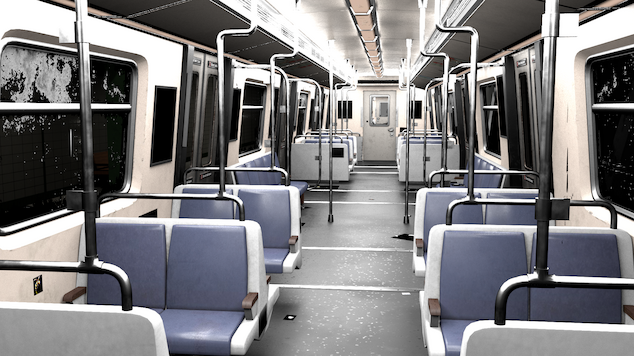} &
\includeimgcroplarge{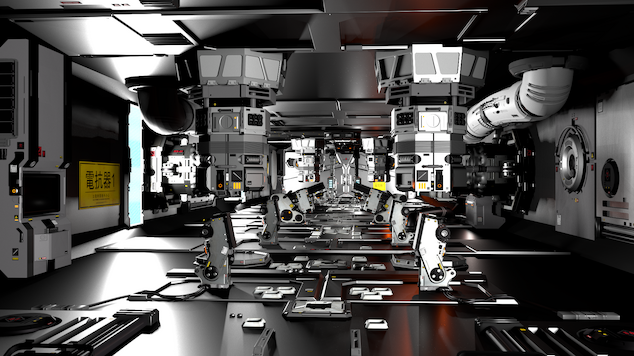} &
\includeimgcroplarge{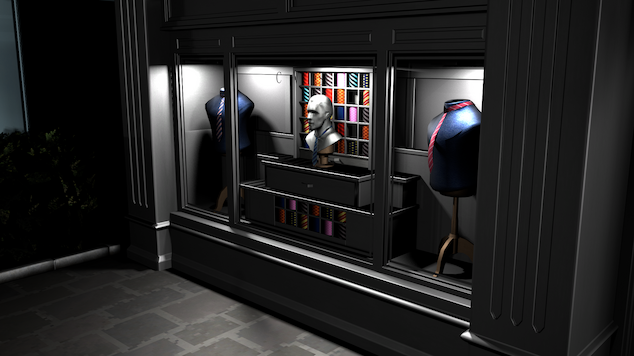} \\
\end{tabular}
\end{minipage}
\end{minipage}
}
\caption{Comparison of tested methods on scenes with many lights, captured after 1024 frames. The numbers under each image correspond to the FLIP error against the ground truth and a total frame time. Note that Cluster NVC is more expensive than NVC compared to ReSTIR, due to $6\times$ more training data needed and a more complex sampling process. For scenes with up to 32 lights, the performance of NVC is comparable to ReSTIR. In these tests, we ran Clustered NVC (C-NVC) to generate initial ReSTIR candidates for every pixel, instead of only on disocclusions as we recommend.}
\label{fig:many_light_scenes}
\end{figure*}

Though using NVC for initial candidates improves the quality of ReSTIR overall, the biggest benefit is for boosting the convergence rate for disocclusions. Figure~\ref{fig:cluster_restir} shows a significant noise reduction when NVC is used for initial candidates after a disocclusion event. Running NVC for ReSTIR once it converges only improves ReSTIR quality insignificantly; the average FLIP reduction is about $5\%$ (see Figure~\ref{fig:many_light_scenes}, while introducing an overhead of running the inference for every pixel. Therefore, we recommend running NVC only for disoccluded pixels to boost the quality of initial candidates. This implementation leads to runtime performance being only $5\%$ lower compared to standard ReSTIR, on a test using a dynamic camera flying around the scene for 1024 frames. Results of clustered NVC in comparison to other methods are summarized in Figure~\ref{fig:many_light_scenes}.

\subsection{Comparison to Neural Radiance Cache}
\label{sec:Comparison_NRC}

\begin{figure*}
\centering
\begin{minipage}{0.33\textwidth}
    \begin{tikzpicture}
        \node[anchor=south west, inner sep=0] (image) at (0,0) {\includegraphics[width=\linewidth]{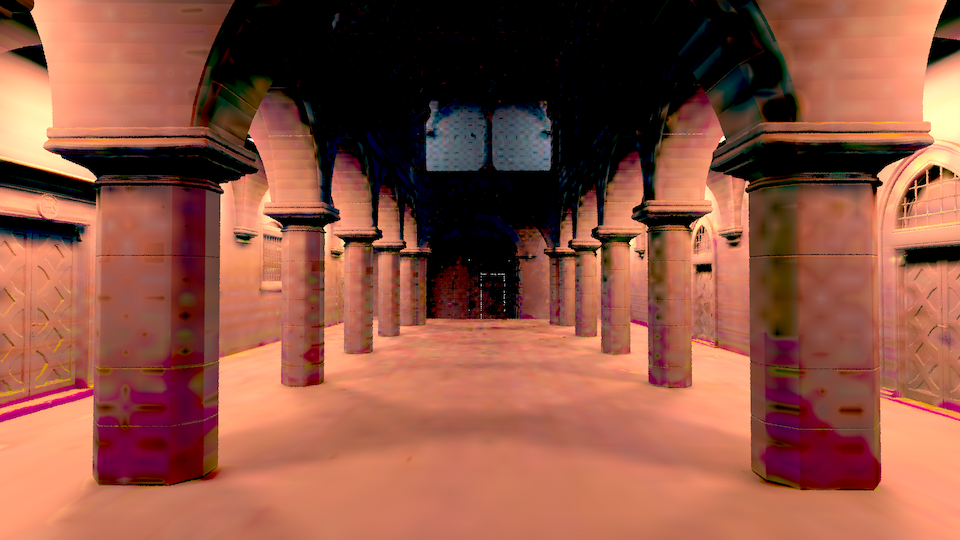}};
        \node[anchor=north west, xshift=1mm, yshift=-1mm, fill=white, opacity=0.7, text opacity=1, font=\tiny] 
            at (image.north west) {NRC};
    \end{tikzpicture}
\end{minipage}%
\begin{minipage}{0.33\textwidth}
    \begin{tikzpicture}
        \node[anchor=south west, inner sep=0] (image) at (0,0) {\includegraphics[width=\linewidth]{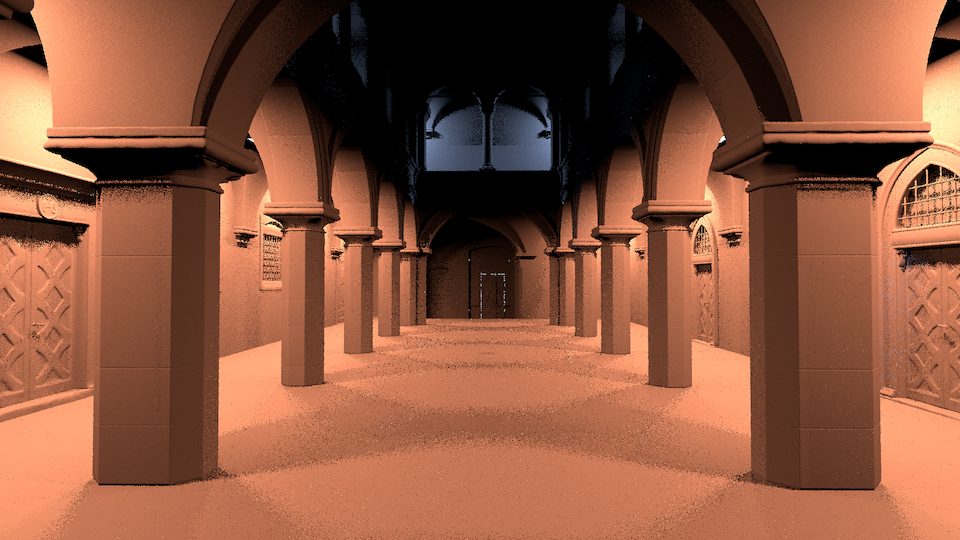}};
        \node[anchor=north west, xshift=1mm, yshift=-1mm, fill=white, opacity=0.7, text opacity=1, font=\tiny] 
            at (image.north west) {Neural Light Sampling (ours)};
    \end{tikzpicture}
\end{minipage}%
\begin{minipage}{0.33\textwidth}
    \begin{tikzpicture}
        \node[anchor=south west, inner sep=0] (image) at (0,0) {\includegraphics[width=\linewidth]{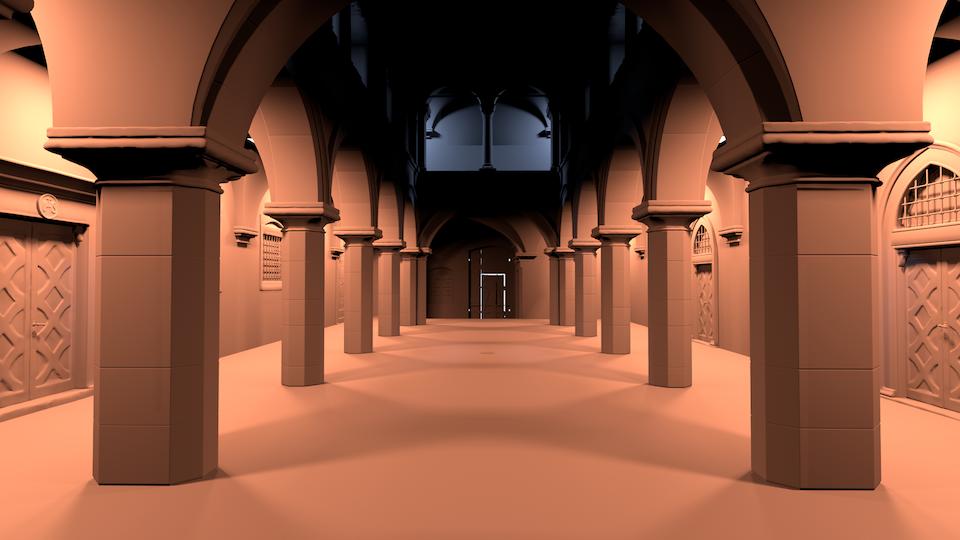}};
        \node[anchor=north west, xshift=1mm, yshift=-1mm, fill=white, opacity=0.7, text opacity=1, font=\tiny] 
            at (image.north west) {Ground Truth};
    \end{tikzpicture}
\end{minipage}%

\caption{\label{fig:nrc_comparison}
 NRC approximating a product of radiance and visibility exhibits blurry artifacts and color shifts (left). Our method computes exact radiance analytically and samples the light sources stochastically for unbiased rendering (middle). The ground truth was accumulated with many frames (right). 
}
\vspace*{-4pt}
\end{figure*}

For a direct comparison of our method to the neural radiance cache (NRC)~\cite{muller2021real}, we have implemented a version of NRC that caches only direct illumination from all lights on a primary hit (NRC DI). We use the same network architecture and training data procedure as we do for our method, except the output layer has three neurons for RGB radiance values, and we use the leaky ReLU activation function to allow for unbounded values. NRC is not limited by the number of lights, but it has several drawbacks compared to our method. For optimal results, NRC implementation requires a larger MLP, which increases the training and inference time. Another disadvantage is that NRC requires more training steps to achieve usable results (see Figure~\ref{fig:training_steps}). At the beginning of training, NRC produces random colors, introducing significant error to the resulting image. NRC predicts a product of the wavelength-dependent radiance and visibility, manifesting blurry artifacts and color shifts (see Figure~\ref{fig:nrc_comparison}). Our method instead calculates the radiance analytically with LTC. The predicted visibility is used to guide the light sampling process, which remains unbiased, even when the network training has not yet converged. In this case, we get an increase in variance instead of bias. Compared to NRC, our method is limited to direct illumination and a smaller number of lights, but for this purpose it achieves unbiased results with a smaller network and does not suffer from artifacts due to under-training and radiance approximation.

\enlargethispage{6pt}
\vspace*{-4pt}
\subsection{Performance}
\label{sec:Performance}

The performance of all tested methods is summarized in Table~\ref{tab:perf_breakdown}. Training data generation with our default configuration (8196~training examples and 32~lights) takes 0.34~ms for the Kitchen scene and 0.43~ms for the Sponza scene. The neural network training step (backpropagation and optimization) takes $\sim0.75$~ms for both scenes. When training data generation and the training itself are implemented as separate steps, where the first step writes out the data into memory for the second step to consume, it is possible to perform multiple training iterations over the same data. As we only perform one step per frame, we can fuse data generation and training together, achieving about $5\%$ speedup. We run inference once per pixel at $\sim1.8$~ms per frame for $1920\times1080$ resolution with 32 lights. The inference itself takes 1.32~ms, while the remaining time is spent on the WRS algorithm. For comparison, replacing the inference call of our NVC by casting a shadow ray toward each of the light sources for every pixel on screen takes about 7~ms for the Kitchen scene with 32 lights (more than $3\times$ higher).

\begin{table}
\smallskip{}
    \footnotesize
    \centering
    \begin{tabular}{cccccc}
    &  G-Buffer & Training (fused) &  Light Sampling & Shading & Frame Total \\
    &  [ms]     &   [ms]   &       [ms]      &  [ms]   &    [ms]  \\                   
    \hline 
    \hline 
    
    \hline 
        \multicolumn{6}{c}{RIS}\\
    \hline                 
    \hline
       Subway    & 1.09   &  -  & 0.62 & 1.16 & 2.92\\
       ZeroDay   & 0.61   &  -  & 0.36 & 0.99 & 2.00\\
       Bistro    & 0.78   &  -  & 0.31 & 1.61 & 2.74\\
       Kitchen   & 0.58   &  -  & 0.47 & 0.57 & 1.65\\
       Sponza    & 1.00   &  -  & 0.27 & 0.67 & 1.97\\
    \hline
    \hline 

    \hline 
        \multicolumn{6}{c}{ReSTIR}\\
    \hline                 
    \hline
       Subway    & 1.17   &  -  & 2.01 & 1.11 & 4.48\\
       ZeroDay   & 0.64   &  -  & 1.67 & 0.92 & 3.43\\
       Bistro    & 0.83   &  -  & 2.30 & 1.83 & 5.15\\ 
       Kitchen   & 0.58   &  -  & 2.45 & 0.53 & 3.61\\
       Sponza    & 1.00   &  -  & 2.23 & 0.64 & 3.95\\
    \hline
    \hline  

    \hline 
        \multicolumn{6}{c}{NVC (ours)}\\
    \hline                 
    \hline
       Kitchen   & 0.60   &  0.99  & 1.74 & 0.52 & 3.95\\
       Sponza    & 0.99   &  1.18  & 1.85 & 0.61 & 4.69\\
    \hline
    \hline  
    
    \hline 
        \multicolumn{6}{c}{Clustered NVC (ours)}\\
    \hline                 
    \hline
       Subway    & 1.11   &  3.01  & 1.87 & 1.11 & 7.37\\
       ZeroDay   & 0.63   &  2.63  & 1.70 & 0.95 & 6.18\\
       Bistro    & 0.80   &  3.20  & 1.71 & 1.87 & 7.85\\   
    \hline
    \hline

    \hline 
        \multicolumn{6}{c}{Clustered NVC$\rightarrow$ReSTIR (ours)}\\
    \hline                 
    \hline
       Subway    & 1.12   &  3.00  & 3.61 & 1.05 & 8.99\\
       ZeroDay   & 0.64   &  2.58  & 3.43 & 0.90 & 7.75\\
       Bistro    & 0.80   &  3.38  & 2.24 & 1.61 & 8.24\\   
    \hline
    \hline  
    
    \hline
    \end{tabular}
    \caption{Performance breakdown of tested methods. Scenes with 32 lights are tested on NVC and many-light scenes on Clustered NVC. Training data generation and network training has been fused into a single pass for better performance. The light sampling column contains time spent in either the resampling loop (RIS), inference and light selection process (NVC and Clustered NVC), all ReSTIR passes (ReSTIR), or ReSTIR passes with Clustered NVC as the initial samples generator (Clustered NVC$\rightarrow$ReSTIR).} 
    \label{tab:perf_breakdown}
\end{table}

\section{Conclusion and Future Work}
\label{sec:conclusions}
We proposed a lightweight neural-based sampling method for real-time direct illumination based on caching the nonbinary visibility. With a minor modification, our method provides unbiased estimates with lower error than screen-space ReSTIR at a similar cost. Compared to ReSTIR, our method operates in the world space, making it more robust to visibility changes with which ReSTIR struggles. In fact, our neural light sampling could be used in combination with ReSTIR to sample the initial candidates. As a world-space method, it can also be used for direct illumination of volumes of participating media. We used a clustered approach to support an arbitrary number of lights with a fixed-size neural network. We also proposed a biased variant that directly uses the visibility estimates, decreasing variance even further at the cost of some bias. Thanks to continuous online training, our method adapts to the dynamic content including animated lights, geometry, and camera. Compared to neural importance sampling of many lights~\cite{figueiredo2025neural}, we are only caching visibility with a smaller MLP, achieving faster training and simpler implementation (we do not require other inputs, e.g., the surface normal to the MLP, only the position).

The limitation of our clustered NVC approach is that efficiency is highly dependent on the quality of the clusters created and the total number of lights. For future work, an interesting direction would be to explore other methods for light clustering, e.g., the ones based on hierarchies~\cite{Vevoda2018BOR, figueiredo2025neural} and light culling~\cite{tokuyoshi2016stochastic}. An early approach of \citet{shirley1996monte} of sorting lights into sets of important and unimportant lights could be adapted to our approach by creating one cluster of unimportant lights to ensure unbiasedness and using the remaining available clusters to represent the important lights. 

Another interesting research direction would be to cache mutual visibility between any two points in the scene. This would enable us to query the visibility of any light, not just the ones represented by output neurons. To make this practical, such a query would either have to be faster than a ray cast or have to return visibility for a batch of queries in a single inference call, to amortize the cost. 

Finally, we need to overcome the limitation of the count of lights (or light clusters) due to the restrictions imposed on the size of the neural network. We can achieve linear scaling na\"{i}vely by having multiple networks and possibly time-slicing their training (only train one network per frame) to maintain a fixed training budget. Overcoming this limitation in a scalable way is a nontrivial task for the future.

\section*{Acknowledgements}
We would like to thank Holger Gr\"un and Carsten Benthin for useful discussions and their support. The models are courtesy of Crytek (Sponza), Sketchfab/Alex Murias (Subway), Blendswap/Jay-Artist (Country Kitchen), Mike Winkelmann (ZeroDay), and Amazon Lumberyard (Bistro).

\small
\bibliographystyle{jcgt}
\bibliography{main}

\section*{Index of Supplemental Materials}
\begin{itemize}
\item Video of Sponza walkthrough\newline
\href{https://jcgt.org/published/0014/02/01/NVC_Sponza_Walkthrough.mp4}{jcgt.org/published/0014/02/01/NVC\_Sponza\_Walkthrough.mp4}

\item Video of Sponza with dynamic lighting\newline
\href{https://jcgt.org/published/0014/02/01/NVC_Sponza_Dynamic_Light.avi}{jcgt.org/published/0014/02/01/NVC\_Sponza\_Dynamic\_Light.avi}
\end{itemize}

\section*{Author Contact Information}

\hspace{-2mm}\begin{tabular}{p{0.5\textwidth}p{0.5\textwidth}}
Jakub Bok\v{s}ansk\'y \newline
Advanced Micro Devices GmbH \newline
Einsteinring 22-28 \newline
85609 Aschheim \newline
Munich, Germany \newline
\href{mailto:jakub.boksansky@amd.com}{jakub.boksansky@amd.com}
\href{https://boksajak.github.io}{https://boksajak.github.io}
&

Daniel Meister \newline
AMD Japan Co. Ltd. \newline
Marunouchi Trust Tower \newline
1-8-3 Marunouchi, Chiyoda-ku \newline
Tokyo, Japan \newline
\href{mailto:daniel.meister@amd.com}{daniel.meister@amd.com}
\href{https://meistdan.github.io}{https://meistdan.github.io}

\end{tabular}

\afterdoc

\end{document}